\documentclass[a4paper,manyauthors,nocleardouble,COMPASS]{cernphprep}
\RequirePackage{flushend}
\RequirePackage[numbers,sort&compress]{natbib}
\usepackage[utf8]{inputenc}
\usepackage{mathptmx}
\usepackage{amsmath}
\usepackage{bm}		
\usepackage{appendix}
\usepackage{lineno} 
\usepackage{upgreek}
\usepackage{booktabs}
\usepackage{xspace}
\usepackage{bigstrut}
\usepackage{graphicx}
\usepackage{amsmath}
\usepackage{url}
\usepackage[nobiblatex]{xurl}
\usepackage{color}
\usepackage{rotating}
\usepackage{dutchcal}
\usepackage[dvipsnames]{xcolor}
\usepackage{float}	
\usepackage{tikz}	

\pdfoutput=1    
\DeclareUnicodeCharacter{2212}{\ensuremath{-}}  
\usepackage{orcidlink}
\usepackage{tikz}
\usepackage{hyperref}
\hypersetup{colorlinks,citecolor=blue,filecolor=blue,linkcolor=blue,urlcolor=blue}
\PassOptionsToPackage{hyphens}{url}\usepackage{hyperref}
\usepackage[capitalise]{cleveref}
\crefrangelabelformat{equation}{(#3#1#4--#5#2#6)}

\newcommand{\di}[0]{\mathrm{d}}

\newcommand{\xsecvil}[4]{$#1 \ {} \pm { \ #2} _{\text{stat}} \ {}_{- \ #3}^{+ \ #4}\big\rvert_{\text{sys}}$}

\renewcommand{\vec}[1]{\bm{#1}}

\newcommand{\bitem}{\begin{itemize}}
\newcommand{\eitem}{\end{itemize}}
\newcommand{\be}{\begin{equation}}
\newcommand{\ee}{\end{equation}}
\newcommand{\bea}{\begin{eqnarray}}
\newcommand{\eea}{\end{eqnarray}}

\newcommand{\xBj}[0]{x_{\rm Bj}}

\newcommand{\mr}[1]{\mathrm{#1}}

\newcommand{\GeVc}{GeV/$c$\,}
\newcommand{\GeVcc}{GeV/$c^{\rm\, 2}$\,}
\newcommand{\GeVcsq}{(GeV/$c$)$^{\rm\, 2}$\,}
\newcommand{\GeVccsq}{(GeV/$c^{\rm\, 2}$)$^{\rm\, 2}$\,}

\newcommand{\qsqU}{$({\rm GeV}\kern -0.15em /\kern -0.08em c)^2$\xspace}

\newcommand{\beq}{\begin{equation}}
\newcommand{\eeq}{\end{equation}}

\renewcommand{\vec}[1]{\bm{#1}}

\renewcommand{\Re}{\mr{Re}}
\renewcommand{\Im}{\mr{Im}}

\newcommand{\pT}{\vec{p}_{\kern -0.15em \perp}}
\newcommand{\pTsq}{p_{\kern -0.15em \perp}^2}
\newcommand{\LeftRight}{\leftrightarrows}
\newcommand{\sigmaT}{\sigma_\mr{T}}
\newcommand{\sigmaL}{\sigma_\mr{L}}
\newcommand{\sigmaTT}{\sigma_\mr{TT}}
\newcommand{\sigmaLT}{\sigma_\mr{LT}}
\newcommand{\sigmaLTp}{\sigma_\mr{LT'}}
\newcommand{\fm}{$\phantom{-}$}
\newcommand{\fz}{$\phantom{0}$}
\newcommand{\phibi}{\small{\fm$-\pi$ -- $\dfrac{-3\pi}{4}$}}
\newcommand{\phibii}{\small{$\dfrac{-3\pi}{4}$ -- $\dfrac{-\pi}{2}$}}
\newcommand{\phibiii}{\small{\fz$\dfrac{-\pi}{2}$ -- $\dfrac{-\pi}{4}$}}
\newcommand{\phibiv}{\small{\fz$\dfrac{-\pi}{4}$ -- $0$}}
\newcommand{\phibv}{\small{\fm$\phantom{\dfrac{\pi}{2}}$$0$ -- $\dfrac{\pi}{4}$}}
\newcommand{\phibvi}{\small{\fm\fz$\dfrac{\pi}{4}$ -- $\dfrac{\pi}{2}$}}
\newcommand{\phibvii}{\small{\fm\fz$\dfrac{\pi}{2}$ -- $\dfrac{3\pi}{4}$}}
\newcommand{\phibviii}{\small{\fm$\dfrac{3\pi}{4}$ -- $\pi$}}

\begin{document}
\begin{titlepage}
\PHnumber{2024-xxx}
\PHdate{\today}
\DEFCOL{CDS-Library}

\title{Measurement of the hard exclusive $\uppi^{0}$ muoproduction cross section at COMPASS}
\Collaboration{The COMPASS Collaboration}
\ShortAuthor{The COMPASS Collaboration}

\begin{abstract}
A new and detailed measurement of the cross section for hard exclusive neutral-pion muoproduction on the proton was performed in a wide kinematic region, with the photon virtuality $Q^2$ ranging from 1 to 8~\GeVcsq and the Bjorken variable $\xBj$ ranging from 0.02 to 0.45. The data were collected
at COMPASS at CERN  using 160~\GeVc longitudinally polarised $\upmu^+$ and $\upmu^-$ beams scattering off a 2.5~m long liquid hydrogen target.
From the average of the measured  $\upmu^+$ and $\upmu^-$ cross sections,  the virtual-photon--proton cross section is determined as a function of the squared four-momentum transfer between the initial and final state proton in the range 0.08~\GeVcsq $< |t| <$ 0.64~\GeVcsq.
From its angular distribution, the combined contribution of transversely and longitudinally polarised photons are determined, as well as transverse--transverse  and longitudinal--transverse interference contributions. They are studied as functions of four-momentum transfer $|t|$, photon virtuality  $Q^2$ and virtual-photon energy $\nu$.
The longitudinal--transverse interference contribution is found to be compatible with zero. The significant transverse--transverse interference contribution reveals the existence of a dominant contribution by transversely polarized photons. This provides clear experimental evidence for the chiral-odd GPD $\overline{E}_T$.
In addition, the existence  of a non-negligible contribution of longitudinally polarized photons is suggested by the $|t|$-dependence of the cross section at  $\xBj < $ 0.1 .
Altogether, these results provide valuable input for future modelling of GPDs and thus of cross sections for exclusive pseudo-scalar meson production. Furthermore, they can be expected to facilitate the study of next-to-leading order corrections and higher-twist contributions.
\end{abstract}
\Submitted{(to be submitted to Physics Letters B)}

\end{titlepage}

\section{Introduction }

Generalized parton distributions (GPDs),
as introduced in Refs~\cite{Muller:1994ses, Ji:1996ek, Ji:1996nm, Radyushkin:1996nd, Radyushkin:1997ki}, are non-perturbative objects, which describe the  three-dimensional structure of the nucleon by correlating transverse spatial positions and longitudinal momentum fractions of the partons (quarks and gluons) of the nucleon. The GPDs contain also rich information about spin and angular momentum  at parton level.
They can be accessed by hard exclusive  pseudoscalar meson leptoproduction on the nucleon, for which the  leading mechanism is presented in Fig.\,1.
Collinear factorization theorems \cite{Collins:1996fb} can be applied  to longitudinally polarised virtual photons and establish that the pseudoscalar meson production amplitude  factorizes into a hard perturbative part and soft components described by the chiral-even GPDs $\widetilde{H}$, $\widetilde{E}$ of the nucleon and
the twist-2 part of the meson wave function.
Contributions from transversely polarised virtual photons to the production of pseudoscalar
mesons are expected to be suppressed in the production amplitude by $1/Q$~\cite{Collins:1996fb}, where $Q^2$ is the virtuality of the photon
that is exchanged between lepton and proton. However,
experimental data on exclusive $\uppi^+$ production from HERMES~\cite{HERMES:2007hrc}
and JLab CLAS~\cite{CLAS:2016tqs, CLAS:2022iqy}
and on exclusive $\uppi^0$
production from JLab CLAS~\cite{CLAS:2007jvi, CLAS:2012cna, CLAS:2014jpc, CLAS:2023wda} and Hall A~\cite{Fuchey:2010xsx, JeffersonLabHallA:2016wye, JeffersonLabHallA:2017hky, JeffersonLabHallA:2020dhq} suggest that such contributions are substantial.
In the collinear approximation, singularities occur for
transversely polarised virtual photons and mesons. Regularization is accomplished in the framework of phenomenological models as in Refs~\cite{Goloskokov:2009ia, Goloskokov:2011rd, Ahmad:2008hp, Goloskokov:2022rtb, Duplancic:2023xrt}  by including transverse degrees of freedom of quarks and antiquarks that make up the meson.
In these models, such transversely polarised virtual-photon contributions are possible
provided a quark helicity-flip GPD couples to a twist-3 helicity-flip meson wave function  and thus pseudoscalar meson production involves also the chiral-odd (transversity) GPDs
$H_\mr{T}$ and $\overline{E}_{\mr{T}}$.

\vspace{0mm}
\setlength\intextsep{20pt}
\begin{figure}[h!]
\centering
\includegraphics[width=0.49\textwidth]{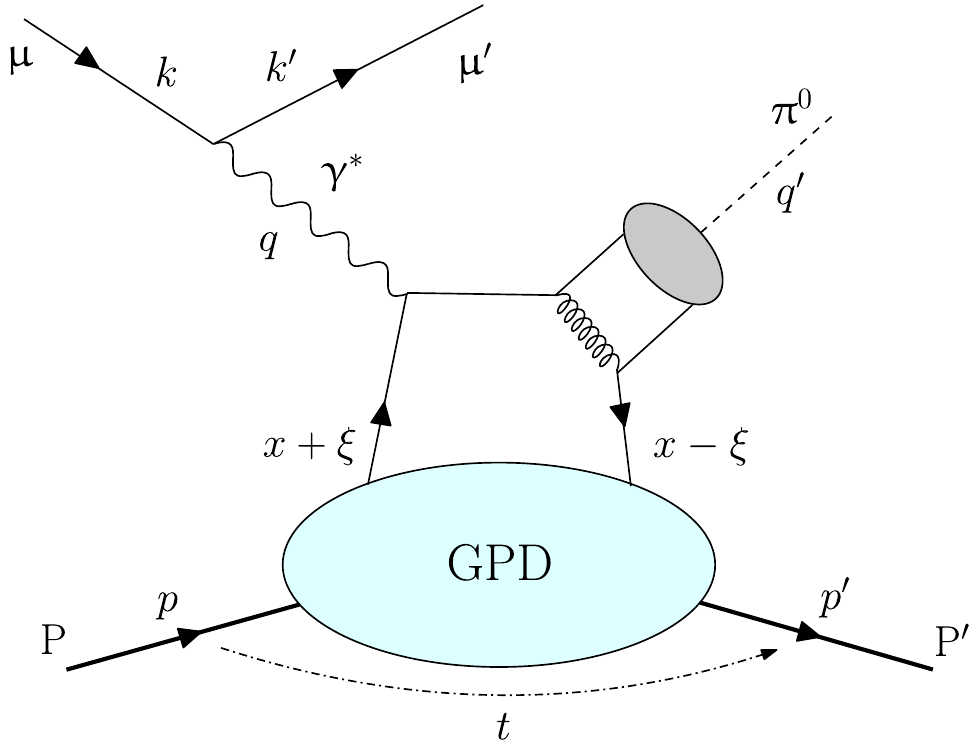}
\caption{\label{fig:handbag}
Leading-twist diagram for hard exclusive $\uppi^0$ leptoproduction off the proton.
Here, $k$, $k'$, $q$, $q'$, $p$, $p'$ are the four-momenta of incident and outgoing muon, virtual photon, outgoing $\pi^0$ and of incident and outgoing proton.
The squared four-momentum transfer between initial and final proton is denoted by $t$, the average longitudinal momentum fraction of the active quark by $x$ and half of the transferred longitudinal momentum fraction by $\xi$.
}
\end{figure}

The GPDs depend on the average longitudinal momentum fraction of the active quark $x$, half of the transferred longitudinal momentum fraction $\xi$,  the squared four-momentum transfer between the initial and final proton $t$ and the virtual photon virtuality $Q^2=-q^2$, see Fig. 1. The
chiral-even GPD $\widetilde{H}$ is related in the forward limit at $t=0$ to the quark helicity distribution $\Delta q(x)$ measured in deep inelastic scattering (DIS), while the chiral-even GPD $\widetilde{E}$ has no such counterpart in DIS. Pion-pole exchange  is expected to give the main contribution to $\widetilde{E}$  at small $t$ for $\uppi^+$ production, while for $\uppi^0$ production the pion pole is absent.
The chiral-odd GPDs,
$H_\mr{T}$ and $\overline{E}_{\mr{T}}$, which are well described in Refs.~\cite{Diehl:2001pm, Diehl:2005jf, Burkardt:2005hp, Burkardt:2006ev}, are related to the transversity and the Boer--Mulders parton distribution functions measured in semi-inclusive DIS. The GPD
 $H_\mr{T}$ describes the correlation
 between the transverse polarisations of quark and proton. It has been shown that in impact-parameter space
$\overline{E}_{\mr{T}}$ is related to a sideways shift in the distribution of transversely polarised quarks in an unpolarised proton. It can be used to determine the correlation between transverse quark spin and quark angular momentum and is connected to the  transverse anomalous magnetic moment of the proton at $t=0$. In contrast to $\uppi^+$ production, $\uppi^0$ production is dominated by $\overline{E}_{\mr{T}}$
due to the quark flavour decomposition of pions and the relative signs of the GPDs for $u$ and $d$ quarks,
while the contributions from $\widetilde{H}$, $\widetilde{E}$ and $H_\mr{T}$ are significantly reduced.

In this paper, we present results for exclusive $\uppi^0$ production in muon--proton scattering,
$\upmu \mr{p} \rightarrow \upmu' \mr{p}' \uppi^0$, which complement the first COMPASS results published~\cite{COMPASS:2019fea, Matthias-Gorzellik} with a comprehensive analysis of a new set of data
in an extended
kinematic domain, thus providing novel input for GPD models, in particular for the chiral-odd GPDs.


\section{Formalism }
\label{sec:theory}

The  cross section for hard exclusive $\uppi^{0}$ production by scattering a polarised muon off an unpolarised proton is {given} as:
\begin{equation}
\label{eq:muon-cross-section}
\frac{\di^4 \sigma_{\upmu \mr{p} }^{\LeftRight}}{\di Q^2 \di \nu \di t \di \phi}
=
\Gamma (Q^2,\nu,E_{\upmu}) \;\;
\frac{\di^2 \sigma _{\upgamma^* \mr{p} }^{\LeftRight}}{ \di t \di \phi}\,,
\end{equation}
where $ \Gamma(Q^2,\nu,E_{\upmu})$ is the transverse-virtual-photon flux~\cite{Hand:1963bb} for muon energy $E_{\upmu}$ and virtual-photon energy  $\nu$ in the proton-target rest frame.
%
The virtual-photon{--}proton cross section reads: 
\begin{align}
\label{eq:xsection}
\frac{\di^2 \sigma _{\upgamma^* \mr{p} }^{\LeftRight}}{ \di t \di \phi}  = \frac{1}{2\pi}
\Bigg[&\epsilon \frac{\di \sigma_{\mr{L}}}{\di t}
+ \frac{\di \sigmaT}{\di t}
+ \epsilon \cos \left( 2\phi \right) \frac{\di \sigmaTT}{\di t}\\
\nonumber
&+ \sqrt{2 \epsilon \left( 1 + \epsilon \right)} \cos \phi
\frac{\di \sigmaLT}{\di t}
\mp |P_l| \sqrt{2 \epsilon(1-\epsilon)} \sin{\phi} \frac{\di \sigmaLTp}{\di t}\Bigg]. &
\end{align}
Here, $\phi$ is the angle between lepton-scattering and $\uppi^0$ production planes following the Trento convention~\cite{Bacchetta:2004jz}. The virtual-photon polarisation parameter
$\epsilon$ is defined as:
\begin{equation}
\epsilon=\frac{1-y-\frac{y^2\gamma^2}{4}}{1-y+\frac{y^2}{2}+\frac{y^2\gamma^2}{4}}, \quad
y=\frac{\nu}{E_\upmu},
\quad \gamma = \frac{Q^{2}}{\nu^{2}},
\end{equation}
and $\sigmaL, \sigmaT, \sigmaTT$, $\sigmaLT$ and $\sigmaLTp$ are cross sections. Here, the subscripts T and L denote the contributions of transversely and longitudinally polarised virtual photons, respectively, {and} the subscripts~TT, LT and LT$'$ denote interference terms.
The sign $\mp$ of the lepton beam polarisation $P_l$ corresponds to the negative and positive helicity of the incoming $\upmu^{+}$ and $\upmu^{-}$, respectively, which is denoted by $\LeftRight$.
The spin-independent cross section can be obtained by averaging the two spin-dependent cross sections as:
\begin{equation}
\label{eq:sum_x_sec}
\frac{\di^2 \sigma _{\upgamma^* \mr{p} }}{ \di t \di \phi}  = \frac{1}{2}\Biggl(\frac{\di^2 \sigma _{\upgamma^* \mr{p} } ^{\leftarrow} }{ \di t \di \phi}+\frac{\di^2 \sigma _{\upgamma^* \mr{p} }  ^{\rightarrow}}{ \di t \di \phi}\Biggr).
\end{equation}

Using $\upmu^{+}$ and $\upmu^{-}$ beams of exactly opposite polarisation, the last term in Eq.~(\ref{eq:xsection}) cancels upon averaging, which leads to the expression:
\begin{equation}
\label{eq:xsection_unpol}
	\frac{\textrm{d}\sigma^{\upgamma^{*}\mr{p}}}{\textrm{d}t\textrm{d}\phi} = \frac{1}{2\pi} \bigg[
 \epsilon \frac{\textrm{d}\sigma_\mr{L}}{\textrm{d}t}
 + \frac{\textrm{d}\sigmaT}{\textrm{d}t}
 + \epsilon \cos(2\phi) \frac{\textrm{d}\sigmaTT}{\textrm{d}t} + \sqrt{2\epsilon(1+\epsilon)}\cos(\phi) \frac{\textrm{d}\sigmaLT}{\textrm{d}t}\bigg].
\end{equation}

The cross sections in Eq.~(\ref{eq:xsection}) and (\ref{eq:xsection_unpol}) are related to convolutions of GPDs and meson wave functions with individual hard scattering amplitudes~\cite{Goloskokov:2009ia}, denoted as $\langle F \rangle$ for a GPD $F$:
\begin{align}
\label{equ::relation_L}
    \frac{\di \sigma_\mr{L}}{\di t} &\propto
    \Big[
        (1-\xi^2) \big|\langle{\widetilde{H}\rangle}\big|^2
        - 2\xi^2 \, \Re\left[ \langle{\widetilde{H}\rangle}^\ast\langle{\widetilde{E}\rangle} \right]
        - \frac{t}{4m_{\rm p}^2}\xi^2\big|\langle{\widetilde{E}\rangle}\big|^2
    \Big],\\
\label{equ::relation_T}
\frac{\di \sigmaT}{\di t} &\propto \Big[ (1-\xi^2) \big|\langle{H_T \rangle}\big|^2 - \frac{t'}{8m_{\rm p}^2}\big|\langle{\overline{E}_\mr{T}\rangle}\big|^2 \Big],\\
\label{equ::relation_TT}
    \frac{\di \sigmaTT}{\di t} &\propto \frac{t'}{16m_{\rm p}^2} \big|\langle{\overline{E}_\mr{T}\rangle}\big|^2,\\
\label{equ::relation_LT}
    \frac{\di \sigmaLT}{\di t} &\propto \xi \sqrt{1-\xi^2}\sqrt{-t'} \;\; \Re\left[ \langle{H_\mr{T}\rangle}^\ast\langle{\widetilde{E}\rangle}
    \right]
     ,\\
\label{equ::relation_LTprime}
    \frac{\di \sigmaLTp}{\di t} &\propto \xi \sqrt{1-\xi^2}\sqrt{-t'} \;\; \Im\left[ \langle{H_\mr{T}\rangle}^\ast\langle{\widetilde{E}\rangle}
    \right].
\end{align}
%
%
Here,
$t' = t - t_\mr{min}$ with $|t_\mr{min}|$ being the kinematically smallest possible value of $|t|$, and $m_{\rm p}$ is the mass of the proton. The quantity $\xi$ defined in the previous section can be approximated at COMPASS kinematics as $\xi \approx \frac{\xBj}{2 - \xBj}$, where $\xBj=Q^2/(2M \nu)$.


\section{Experimental setup }
\label{sec:compass}

The data were collected
during twelve weeks in 2016
using the COMPASS setup, which was a fixed-target experiment located at the M2 beamline of the CERN Super Proton Synchrotron, using naturally polarised muon beams of both charges with energies of 160~\GeVc.

The incoming muons were detected in a beam momentum spectrometer and a beam telescope.
The final-state particles were detected in a two-stage magnetic spectrometer containing a large variety of tracking detectors, electromagnetic and hadron calorimeters, a ring-imaging Cherenkov detector and muon filters for particle identification. The electromagnetic calorimeter ECAL1 was located in the large-angle part of the spectrometer, while ECAL2 was placed in its small-angle part. Detailed descriptions of the setup can be found in Refs.~\cite{COMPASS:2007rjf, COMPASS:2014cka} .

For the measurements to study GPDs, which are described in this paper,
a 2.5~m long liquid-hydrogen target inserted in a recoil-proton detector CAMERA and a~new electromagnetic calorimeter ECAL0
were added to the setup.
The 4~m long recoil-proton detector consisted of two~concentric barrels equipped with 24~scintillator slabs each. It measured time and
distance of flight between the two barrels to determine momentum and angle of the recoil proton. The ECAL0 calorimeter, located directly downstream of the target, allowed the detection of photons emitted at large polar angles, expanding the accessible kinematic domain towards higher values of $\xBj$. Compared to the previous measurement in 2012, the ECAL0  acceptance was enlarged allowing to cover a larger kinematic range than in Ref.~\cite{COMPASS:2019fea}.
The accessible kinematic domain of the COMPASS spectrometer for measuring exclusive events ranges from  $\xBj \sim$~0.02 to 0.45, which is complementary to other experimental facilities.

Data were recorded with both positively and negatively charged muon beams. Due to
the natural polarisation of the muon beams,  which originates from the parity-violating decay-in-flight of the parent mesons, $\upmu^+$ and $\upmu^-$ beams have opposite polarisation. This enables us to measure the spin-independent cross section, see~\cref{eq:sum_x_sec,eq:xsection_unpol}.
The absolute value of the average
polarisation for both beams is about 0.8 with an uncertainty of about 0.04.

In contrast to the four weeks long pilot run in 2012, for the 2016 data
used in this analysis  comparable {beam} intensities of about $4 \cdot 10^6\,\upmu$/s
were used.
The incoming reconstructable muon flux was measured using a random trigger based on a radioactive decay~\cite{Mount:1981ri}. A set of selection criteria was applied to ensure the quality of the muon tracks and to keep intensity variations below 10\%.
The integrated luminosity used in this analysis
is  51.4~$\mathrm{pb}^{-1}$  for the $\upmu^{+}$ beam and  44.5~$\mathrm{pb}^{-1}$  for the $\upmu^{-}$  beam, measured with a precision of 2\%.


\section{Event selection}
\label{sec:event_selection}

Event candidates for exclusive $\uppi^0$ production are selected from data taken in stable beam and spectrometer conditions. An interaction vertex associated with an incoming muon is required, and only one outgoing track of the same charge is allowed. The latter is identified as scattered muon by requiring that it traverses more than 15 radiation lengths in the spectrometer and is compatible with the trigger conditions.
For incoming muons, the same selection criteria are used as in the flux determination.

The $\uppi^0$ is reconstructed using its dominant two-photon decay.
The photons are detected in the electromagnetic calorimeters ECAL0 and ECAL1. The electromagnetic calorimeter ECAL2, placed in the small-angle part of the spectrometer, is not used in this analysis as it does not provide a significant increase in the statistics of the $\uppi^0$ sample.
In ECAL1 only clusters not associated with a reconstructed charged track are used,
while all clusters are included in ECAL0 as there is no tracking system in front of it. Thresholds of 2~GeV (2.5~GeV) for the higher-energy cluster and of  0.5~GeV (0.63~GeV) for the lower-energy cluster in ECAL0 (ECAL1)  are used.
Compared to the previous analysis of the 2012 data, the threshold of the lower energy cluster was increased to get a better signature of the photon in the electromagnetic calorimeters.
Using the information from the interaction vertex and each combination of two clusters, the kinematics of the recoil proton can be predicted from the momentum balance of the reaction
$\upmu \mr{p} \rightarrow \upmu' \mr{p}' \uppi^0$, $\uppi^0 \rightarrow \upgamma \upgamma$.
The properties of the recoil proton predicted by the spectrometer information are compared
to  the properties of each track candidate reconstructed with the CAMERA system assigning the proton mass.

These recoil-proton candidates are examined by using four so-called exclusivity variables in the further selection process. These are i) the difference $\Delta \varphi$
between predicted and measured azimuthal angle of the recoil proton candidate, ii) the
difference $\Delta p_{\text{T}}$ between predicted and measured transverse momentum $p_{\text{T}}$ of the recoil proton candidate with  respect to the beam direction in the target rest frame, iii) the difference $\Delta z$ between predicted and measured hit
position in the inner ring of CAMERA and iv) the undetected mass $M_{\text{X}}^2$ given by:
\begin{equation}
M_{\text{X}}^2=(k+p-k'-p'-p_{\upgamma_1}-p_{\upgamma_2})^2\,.
\end{equation}
Here,
$k$, $k'$,  $p$ and $p'$ are the four-momenta of incident muon, outgoing muon, target proton and recoil proton respectively,
while $p_{\upgamma_1}$ and $p_{\upgamma_2}$ denote the four-momenta of the two produced
photons.
The following constraints are applied to the exclusivity variables:
\begin{eqnarray*}
|\Delta \varphi|      \hspace{-0.2cm}& < &\hspace{-0.2cm} \text{0.4~rad},\\
|\Delta p_{\text{T}}| \hspace{-0.2cm}& < &\hspace{-0.2cm} \text{0.3~\GeVc},\\
|\Delta z|            \hspace{-0.2cm}& < &\hspace{-0.2cm} \text{16~cm},\\
|M^2_X|               \hspace{-0.2cm}& < &\hspace{-0.2cm} \text{0.3~\GeVccsq.}
\end{eqnarray*}
In addition,
the range $$\text{0.1061} < M_{\gamma \gamma}/\text{(\GeVcc)} < \text{0.1605}$$
is selected for the invariant mass $M_{\gamma \gamma}$.
If more than one combination of vertex, cluster pair and recoil-track candidate
exist, which satisfy the aforementioned selection
criteria for a given event, the
event is excluded from further analysis.

\begin{figure}[h!]
\centering
\includegraphics[width=0.49\textwidth]{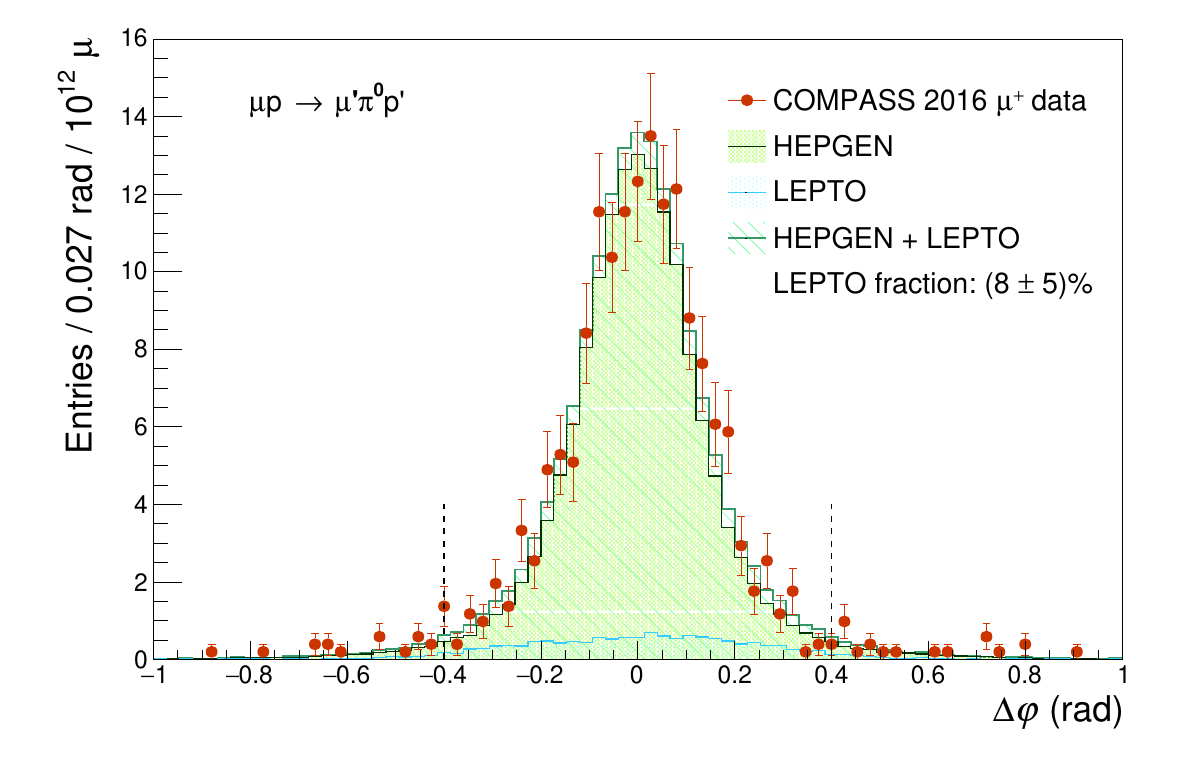}
\includegraphics[width=0.49\textwidth]{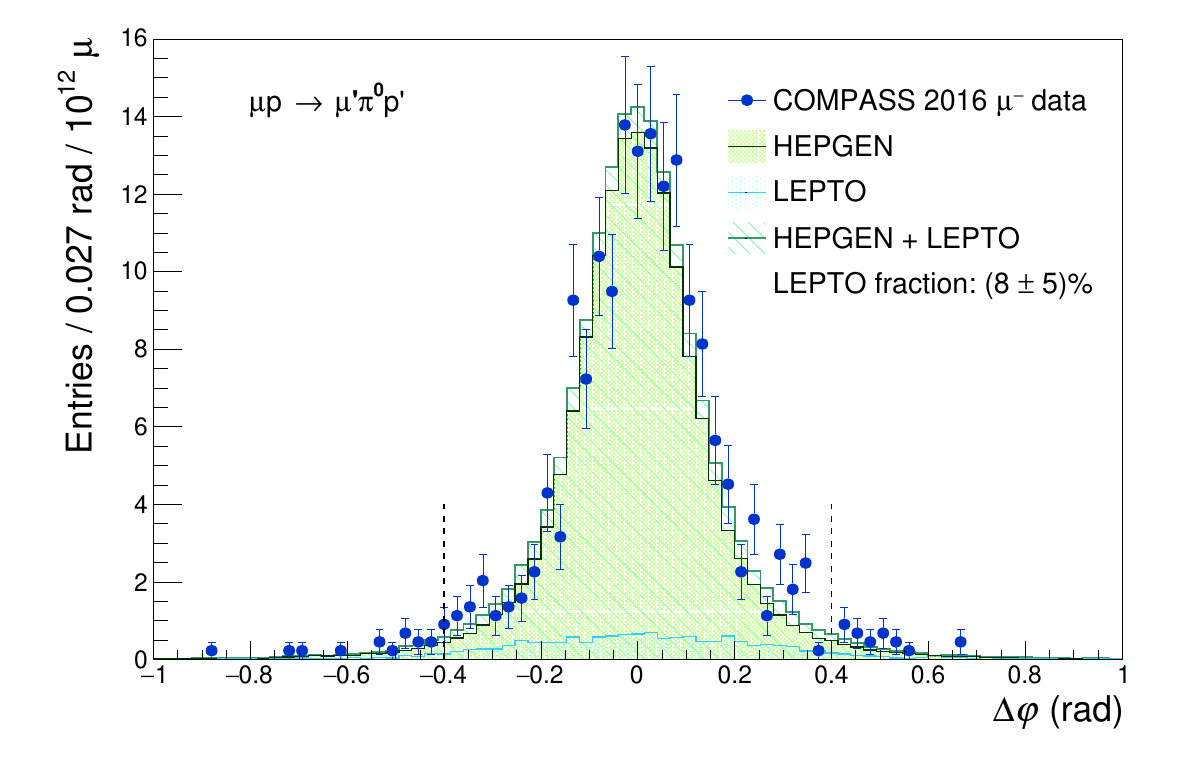}
\includegraphics[width=0.49\textwidth]{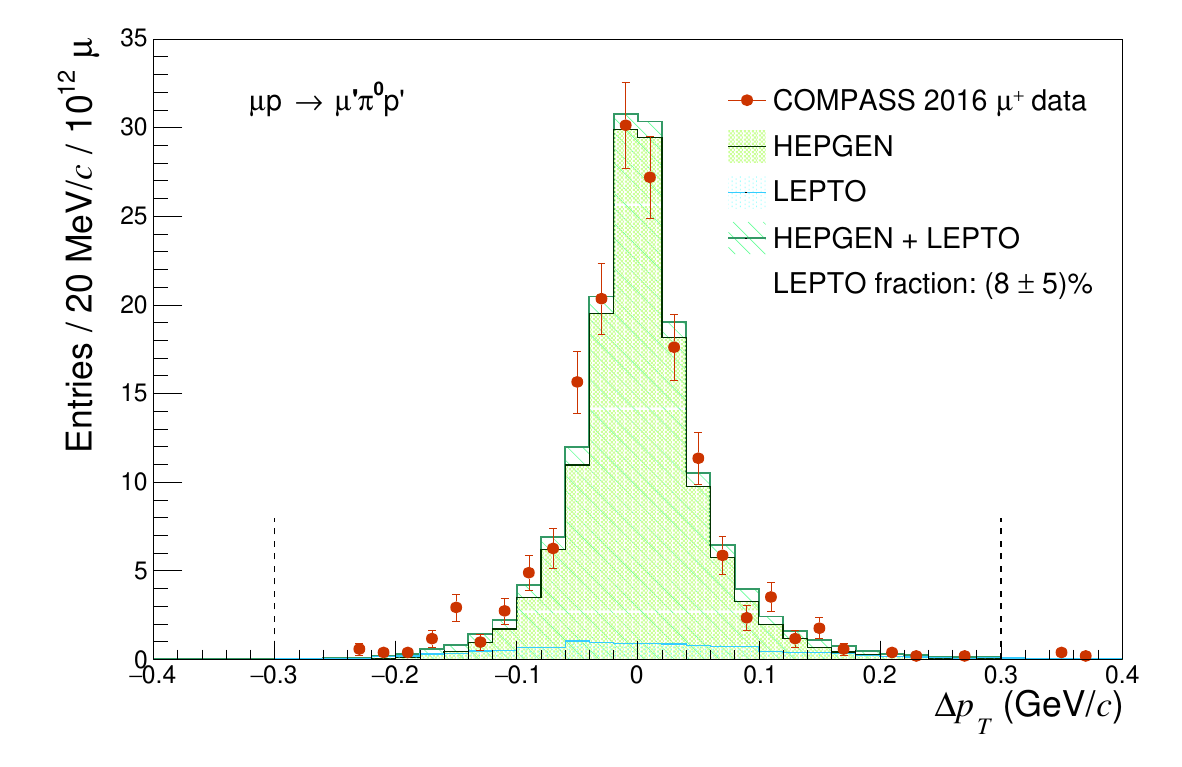}
\includegraphics[width=0.49\textwidth]{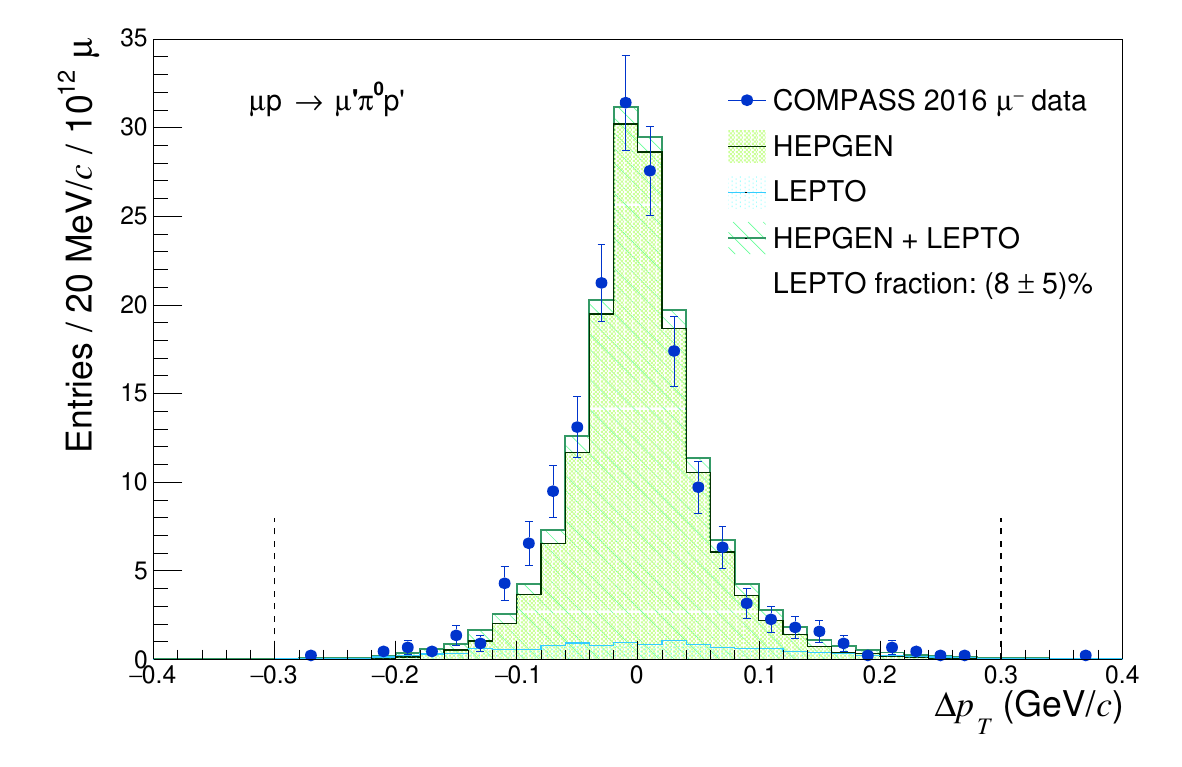}
\caption{\label{fig:selections}
Measured and simulated
distributions $\Delta\varphi$ (top) and $\Delta p_{\text T}$ (bottom), shown for both $\upmu^+$ (left) and $\upmu^-$ (right) beams.
All the distributions are normalised to the same muon flux and the simulations are scaled as described in the text.
The two dashed vertical lines indicate the constraints applied for the selection of events. Error bars denote statistical uncertainties.
The non-exclusive $\uppi^{0}$ background is estimated using LEPTO (blue), while the total $\uppi^{0}$ distribution is expected to match the sum of LEPTO and HEPGEN (dark green).}
\end{figure}

Figure~\ref{fig:selections} shows the $\Delta \varphi$ and $\Delta p_{\text{T}}$ distributions for data obtained with $\upmu^+$ and $\upmu^-$ beams.
They are compared to Monte Carlo (MC) yields of the exclusive (signal) and non-exclusive (background) processes provided by the HEPGEN and LEPTO generators, respectively. This will be explained in more detail in Sect.~\ref{sec:background_estimation}.
A very good agreement between the $\upmu^+$ and $\upmu^-$ data is observed.

\vspace{-1mm}
\setlength\intextsep{20pt}

The purity of the selected events is enhanced by using the over-constrained kinematics from the combined information coming from CAMERA and the spectrometer in a so-called kinematic fit assuming the reaction $\upmu \mr{p} \rightarrow \upmu' \mr{p}' \uppi^0$.
A value smaller than 7 is required for the reduced {$\chi^2$} in order to minimise the non-exclusive background while keeping all events from exclusive
$\uppi^0$ muoproduction.
The kinematic fit also allows to considerably
improve the precision of the event kinematics, in particular for the variables $t$, $Q^2$, $\nu$ and $\phi$.

The data for further analysis are selected within the following kinematic range:
\begin{eqnarray*}
\text{0.08~\GeVcsq}   \hspace{-0.3cm} &< |t| <& \hspace{-0.2cm}\text{0.64~\GeVcsq,} \\
\text{1~\GeVcsq}      \hspace{-0.3cm} &< Q^2 <& \hspace{-0.2cm}\text{8~\GeVcsq,}    \\
\text{6.4~GeV}        \hspace{-0.3cm} &< \nu <& \hspace{-0.2cm}\text{40~GeV.}
\end{eqnarray*}


\section{Background contribution}
\label{sec:background_estimation}

The main source of background to exclusive $\uppi^{0}$ production originates from non-exclusive DIS processes. In order to simulate this background process, the LEPTO 6.5.1 generator with the COMPASS
tuning from Ref.~\cite{COMPASS:2012pfa} is used. Note that events with the topology of exclusive $\uppi^{0}$ production are removed from the LEPTO sample.
For the simulation of signal events, i.e. exclusive $\uppi^{0}$ muoproduction, the HEPGEN++ generator~\cite{HEPGEN,HEPGEN++} is used.
The events from both generators are passed through a complete description of the COMPASS setup~\cite{TGeant} and reconstructed in the same
way as experimental data.

\begin{figure}[h!]
\centering
\includegraphics[width=0.49\textwidth]{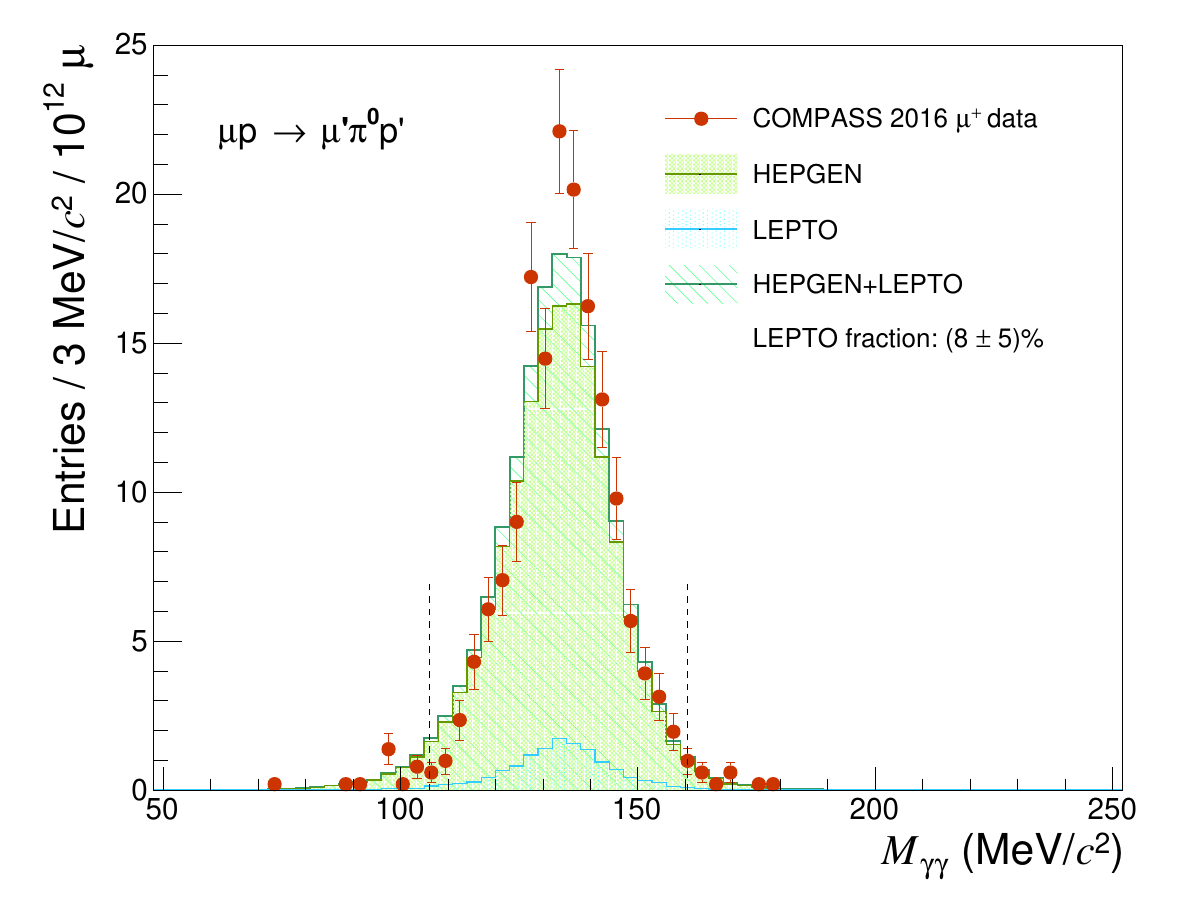}
\includegraphics[width=0.49\textwidth]{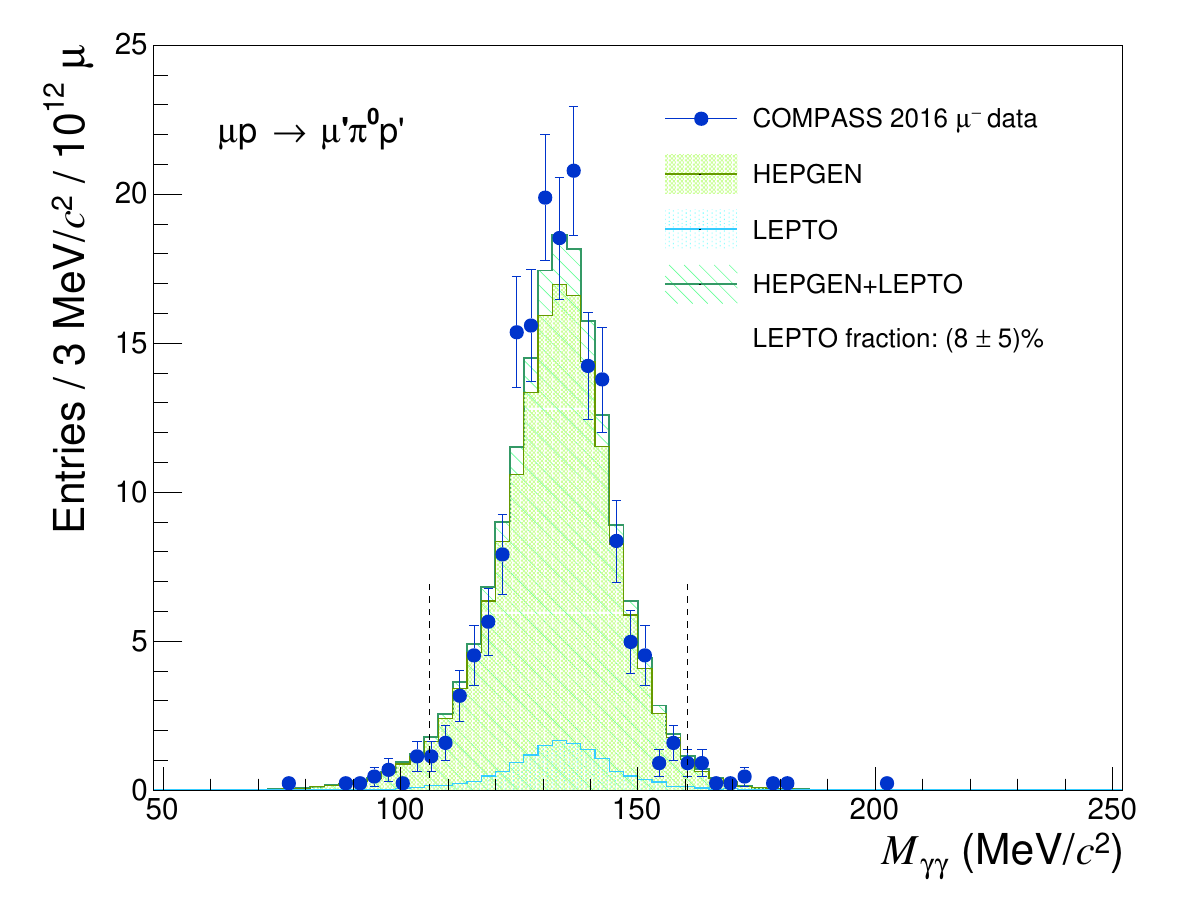}
\caption{\label{fig:mass}
Measured and simulated distributions of $M_{\upgamma\upgamma}$ for data obtained with $\upmu^{+}$ (left) and $\upmu^{-}$ (right) beams.
All the distributions are normalised to the same muon flux and the simulations are scaled as described in the text.
The two dashed vertical lines indicate the interval applied for event selection and normalisation. The non-exclusive $\uppi^{0}$ background is estimated using LEPTO (blue), while the total $\uppi^{0}$ distribution is expected to match the sum of LEPTO and HEPGEN (dark green).}
\end{figure}

The simulated HEPGEN++ and LEPTO Monte Carlo data sets are used to obtain the yields of the exclusive (signal) and non-exclusive (background) processes, respectively. The procedure starts with the determination of a normalisation factor of each MC data set,
$c_{\text L}$ and $c_{\text H}$,
{adjusted} separately to the data in the studied {range in $t$, $Q^2$ and $\nu$} using the distribution of the invariant mass $M_{\upgamma\upgamma}$ around the $\uppi^{0}$ peak. The $M_{\upgamma\upgamma}$ distribution in the full kinematic range is shown in Fig.~\ref{fig:mass}.
Subsequently, the fraction $r_{\text{L}}$
of non-exclusive background
is determined by adjusting the sum of the HEPGEN++ and the LEPTO MC data sets to the experimental data.
For this purpose the variables $\Delta p_\mr{T}$ and $\Delta\varphi$ are used, which are sensitive to the exclusivity of the event. The resulting fraction of non-exclusive background, $r_{\text L}=\left(8 \pm5\right)\%$, is independent on the studied range in $t$, $Q^2$ and $\nu$.
The background determination method {yields} one of the main contributions to the systematic uncertainties. The relative systematic uncertainties of the cross section originating from the background estimation vary from 6 to 16\% depending on the kinematic region.
The scaling factor $f=c_{\text L} \cdot r_{\text L}$ 
is used to correct the data for non-exclusive background,
 see Sec.~\ref{sec:cross_section}.

In comparison to the previous analysis of the  2012 data, the LEPTO fraction was considerably reduced by two reasons. First,
the kinematic fit with a $\chi^2$ requirement has significantly reduced the non-exclusive background, and secondly the HEPGEN++ sample was improved
by reweighting the $\phi$, $t$ and $\nu$ distributions, in order
to obtain a better description of the data.
For the $\phi$ reweighting an iterative process is used to include the  extracted $\phi$ modulation from the data in the model used in HEPGEN++ according to Eq.~(\ref{eq:xsection_unpol}). The reweighting in $t$ and $\nu$ is also applied iteratively using two-dimensional data distributions in these two variables.

A very small background originates from the production of exclusive $\upomega$ mesons with
the decay \mbox{$\upomega \rightarrow \uppi^0 \upgamma$}, where  the photon {is} not detected.
It is estimated by measuring the yield of visible $\upomega$ mesons in the sample, when
combining the exclusive $\uppi^0$ candidate 
with every neutral cluster of energy between the low and high thresholds.
These events are rejected, they represent only  2.7\% of the selected events. They allow the normalization of the HEPGEN++ MC  for exclusive $\upomega$ production. In this way, the contamination of the remaining  $\uppi^0$ sample with not detected $\upomega$ mesons
is evaluated to be about 2.4\% and included in the systematic uncertainties. A possible contribution of the
$\upomega \rightarrow \uppi^0 \uppi^+ \uppi^-$ decay is found to be negligible, as well as from the $\upeta \rightarrow \uppi^0 \uppi^+ \uppi^-$ or $\uppi^0 \uppi^0 \uppi^0$ decay.

{The final data set for the analysis of exclusive
$\uppi^0$ production comprises 1490~events, among which 792~events were collected with the $\upmu^+$ beam and  698~events with the $\upmu^-$ beam}.


\section{Cross section determination and systematics}
\label{sec:cross_section}

The virtual-photon--proton cross section is obtained from the measured muon--proton cross section using:
\begin{equation}
\label{eq:flux_relation}
\frac{\di \sigma^{\upgamma^*\mr{p}}}{ \di |t| \di \phi}= \frac{1}{\Gamma(Q^2,\nu,E_{\upmu})}
\frac{\di  \sigma^{\upmu \mr{p} }}{\di Q^2 \di \nu \di |t| \di \phi},
\end{equation}

\noindent where the transverse-virtual-photon flux $\Gamma(Q^2,\nu,E_{\upmu})$ is given by:

\begin{equation}
\label{eq:photon_flux}
\Gamma(Q^2,\nu,E_{\upmu})  = \frac{\alpha_{\text{em}}  (1- \xBj)}{2 \pi Q^2 y E_{\upmu}} \Bigg[ y^2 \bigg(1 - \frac{2 m_{\upmu}^2}{Q^2} \bigg)
 + \frac{2}{1+Q^2/\nu^2} \bigg(1-y - \frac{Q^2}{4E_{\upmu}^2} \bigg) \Bigg].
\end{equation}

Here, $m_{\upmu}$ denotes the mass of the muon and $\alpha_{\text{em}}$ the electromagnetic fine-structure constant.

The cross section is calculated in a four-dimensional grid with a phase-space element given by $\Delta \Omega_{nijk}= \Delta |t|_n \Delta \phi_i \Delta Q^2_j \Delta \nu_k$.
The binning of the four-dimensional grid is given in the first four columns of Table~\ref{tab:1_2}, while the grid for the variables $Q^2$ and $\nu$ used in the previous publication~\cite{COMPASS:2019fea} is reported in the last two columns for comparison.

\begin{table}[h!]
\caption{\label{tab:1_2}
Four-dimensional grid used for the calculation of the cross section and acceptance. The full {range} for each {variable} is given
in the bottom row of the table.}
\centering
\resizebox{\textwidth}{!}{%
\begin{tabular}{l l l l l l }
\hline \\[-6pt]
               &  & \multicolumn{2}{c}{this work}        & \multicolumn{2}{c}{Ref.~\cite{COMPASS:2019fea}}    \\[2pt]
\hspace{0.5cm}$\phi$ [rad]    & $|t|$ $\bigl[$\GeVcc$\bigr]$  & $Q^2$ $\bigl[$\GeVcc$\bigr]$  & $\nu$ [\text{GeV}]    &  \qquad $ Q^2$ $\bigl[$\GeVcc$\bigr]$ &  $\nu$ [\text{GeV}]  \\[2pt]
\hline \\[-6pt]
\phibi   & 0.08 -- 0.15   & 1.00 -- 1.50 & \fz6.40 -- \fz8.50 & \qquad 1.00 -- 1.50  & \\[8pt]
\phibii  & 0.15 -- 0.22   & 1.50 -- 2.24 & \fz8.50 -- 11.45   & \qquad 1.50 -- 2.24  & \fz8.50 -- 11.45  \\[8pt]
\phibii  & 0.22 -- 0.36   & 2.24 -- 3.34 &  11.45 -- 15.43    & \qquad 2.24 -- 3.34  &  11.45 -- 15.43   \\[8pt]
\phibiv  & 0.36 -- 0.5    & 3.34 -- 5.00 &  15.43 -- 20.78    & \qquad 3.34 -- 5.00  &  15.43 -- 20.78   \\[8pt]
\phibv   & 0.5\fz -- 0.64 & 5.00 -- 8.00 &  20.78 -- 26.00    &                      &  20.78 -- 28.00   \\[8pt]
\phibvi  &                &              &  26.00 -- 40.00    &                      &        \\[8pt]
\phibvii &                &              &                    &                      &        \\[8pt]
\phibviii&                &              &                    &                      &        \\[8pt]
\hline \\[-8pt]
${\Delta} \phi$ [rad]  & ${\Delta}|t|$ $\bigl[$\GeVcc$\bigr]$  & ${\Delta} Q^2$ $\bigl[$\GeVcc$\bigr]$  & ${\Delta} \nu$ [\text{GeV}] & \qquad ${\Delta} Q^2$ $\bigl[$\GeVcc$\bigr]$ & ${\Delta} \nu$ [\text{GeV}] \\[2pt]
\hline \\[-6pt]
$2\pi$      & 0.56 & 7.00     & 33.60    &  \qquad 4.00 & 19.50 \\[2pt]
\hline
\end{tabular}}
\end{table}

 The cross section can be calculated for each beam charge (denoted by $\pm$ in the following) in a bin $\Delta\Omega_{nijk}$ from the data by subtracting the LEPTO MC background:
\begin{equation}\label{Cross_section1}
     \left<\frac{\mr{d}\sigma^{\upmu \mr{p} \rightarrow \upmu' \uppi^0 \mr{p}'}}{\mr{d} \Omega}\right>_{nijk}^{\pm} =
     \left<\frac{\mr{d}\sigma_{\text{data}}^{\upmu \mr{p} \rightarrow \upmu' \uppi^0 \mr{p}'}}{\mr{d} \Omega}\right>_{nijk}^{\pm}
     - \left<\frac{\mr{d}\sigma_{\text{background}}^{\upmu \mr{p} \rightarrow \upmu' \uppi^0 \mr{p}'}}{\mr{d} \Omega}\right>_{nijk}^{\pm}
\end{equation}
The cross section for each four-dimensional bin is calculated as:

\begin{equation}
\label{nijk cross section}
        \left<\frac{\mr{d}\sigma^{\upgamma^*\mr{p} \rightarrow \uppi^0 \mr{p}'}}{\mr{d} |t|\mr{d}\phi}\right>^{\pm}_{nijk} = 
        \frac{1}{\mathcal{L}^{\pm} \Delta t_n \Delta\phi_i \Delta Q^2_j \Delta \nu_k}
         \left(a_{nijk}^{\pm}\right)^{-1}\left( \sum_{l=1}^{N_{nijk}^{ \text{data}}\pm} \frac{1}{\Gamma\left(Q^2_l,\nu_l,E_{\mu \,l}\right)} \right.
        - \left.  f^{\pm}  \cdot \sum_{l=1}^{N_{nijk}^{ \text{L}}\pm} \frac{1}{\Gamma\left(Q^2_l,\nu_l,E_{\mu \,l}\right)} \right),
\end{equation}

where  $\mathcal{L}$ denotes the luminosity and $f$  the normalisation factor that was introduced in Sec.~\ref{sec:background_estimation}. The acceptance denoted as $a_{nijk}$ is determined for each four-dimensional phase-space element using the data simulated by HEPGEN++ as the number of reconstructed events divided by the number of generated events.
Within the phase-space element $\Delta \Omega_{njkl}$,
$N^{\text{data}}_{njkl}$ is the number of measured
events and $N^{\text{L}}_{njkl}$ the number of
LEPTO events.
The virtual-photon flux enters as a kinematic prefactor applied event by event.

The spin-dependent virtual-photon cross section measured with positively or negatively charged muons is obtained in each bin of $\left(|t|,\phi\right)$:
\begin{equation}
     \left<\frac{\mr{d}\sigma^{\upgamma^*\mr{p} \rightarrow \uppi^0 \mr{p}'}}{\mr{d} |t|\mr{d}\phi}\right>^{\pm}_{ni} = \frac{ \sum_{j,k} \left<\frac{\mr{d}\sigma^{\upgamma^*\mr{p} \rightarrow \uppi^0 \mr{p}'}}{\mr{d} |t|\mr{d}\phi}\right>^{\pm}_{nijk} \Delta Q^2_j \Delta\nu_k}{\sum_{j}\Delta Q^2_j \sum_{k} \Delta\nu_k}.
\end{equation}
The spin-independent virtual-photon cross section is calculated as the average of the two spin-dependent cross sections:
\begin{equation}
     \left<\frac{\mr{d}\sigma^{\upgamma^*\mr{p} \rightarrow \uppi^0 \mr{p}'}}{\mr{d} |t|\mr{d}\phi}\right>_{ni} = \frac{1}{2} \left(\left<\frac{\mr{d}\sigma^{\upgamma^*\mr{p} \rightarrow \uppi^0 \mr{p}'}}{\mr{d} |t|\mr{d}\phi}\right>^{+}_{ni}
     + \left<\frac{\mr{d}\sigma^{\upgamma^*\mr{p} \rightarrow \uppi^0 \mr{p}'}}{\mr{d} |t|\mr{d}\phi}\right>^{-}_{ni} \right).
\end{equation}
The cross section can be integrated over the full 2$\pi$-range in $\phi$ in order to study its $|t|$-dependence:
\begin{equation}
     \left<\frac{\mr{d}\sigma^{\upgamma^*\mr{p} \rightarrow \uppi^0 \mr{p}'}}{\mr{d} |t|}\right>_{n} = \sum_{i}\Delta\phi_i \left<\frac{\mr{d}\sigma^{\upgamma^*\mr{p} \rightarrow \uppi^0 \mr{p}'}}{\mr{d} |t|\mr{d}\phi}\right>_{ni}.
\end{equation}
Similar to the study of the $\phi$-modulation of the cross section, the $|t|$-averaged cross section in the measured $\Delta|t|$ range is determined as:
\begin{equation}
     \frac{1}{\Delta|t|}
     \left<\frac{\mr{d}\sigma^{\upgamma^*\mr{p} \rightarrow \uppi^0 \mr{p}'}}{ \mr{d}\phi}\right>_{i} =
     \frac{1}{\sum_{n}\Delta|t|_n}
     \sum_{n}\Delta|t|_n \left<\frac{\mr{d}\sigma^{\upgamma^*\mr{p} \rightarrow \uppi^0 \mr{p}'}}{\mr{d} |t|\mr{d}\phi}\right>_{ni}.
\end{equation}
Using this $\phi$-distribution of the spin-independent cross section,
the contributions to the cross section, i.e. $\frac{\textrm{d}\sigma_\mr{T}}{\textrm{d}t}+\epsilon \frac{\textrm{d}\sigma_\mr{L}}{\textrm{d}t}$,
$\frac{\textrm{d}\sigma_\mr{TT}}    {\textrm{d}t}$
and
$\frac{\textrm{d}\sigma_\mr{LT}}
{\textrm{d}t}$, can be extracted.

{\color{black}
The systematic uncertainties are evaluated separately for each of the results presented below.
Table~{\ref{tab:systematics}}
shows the estimated relative systematic uncertainties on the measured $|t|$ and $\phi$-dependent cross sections
 in the full kinematic range.
The systematic uncertainties
are  arranged in four groups. The first group contains the systematic uncertainties on the determination of the beam flux for both muon charges. The second group contains the systematic effects related to the uncertainties in the acceptance determination and in the energy thresholds for the lower-energy photon in ECAL0 and ECAL1. 
The third group contains the systematic uncertainties related to the selection of pure exclusive $\uppi^{0}$ events. Here, variations of the $\chi^2$ requirement for the kinematic fit from 4 to 10, of the  fraction of non-exclusive background from 3 to 13\%, and of the LEPTO normalisation factor by 20\% are evaluated using the data. The small systematic uncertainty due to undetected $\omega$-production is estimated by Monte Carlo in each kinematic range.
Radiative corrections are not applied, but an estimate of their size was obtained by {A.\,Afanasev~\cite{Afanasev:2023}} using the method discussed in
Ref.~\cite{Afanasev:2002ee}. It depends on the kinematic limits chosen and varies slightly over the $|t|$ and $\phi$-ranges.
The largest systematic effects come from the third group, mainly from the uncertainty related to the estimation of the non-exclusive background as described
in Sect.~\ref{sec:background_estimation}.

The extracted cross-section values
$\frac{\textrm{d}\sigma_\mr{T}}{\textrm{d}t}+\epsilon \frac{\textrm{d}\sigma_\mr{L}}{\textrm{d}t}$,
$\frac{\textrm{d}\sigma_\mr{TT}}    {\textrm{d}t}$ and
$\frac{\textrm{d}\sigma_\mr{LT}}    {\textrm{d}t}$
are affected by the same systematic effects.
The value of
$\frac{\textrm{d}\sigma_\mr{LT}}    {\textrm{d}t}$ is found to be compatible with zero within statistical uncertainties and its systematic uncertainty is estimated to be at least two-third of the absolute value of the systematic uncertainty on $\frac{\textrm{d}\sigma_\mr{TT}}    {\textrm{d}t}$.
The extracted cross-section contributions
could depend on the sin$\phi$ term in the cross section (see Eq.~(\ref{eq:xsection})), if the opposite
$\upmu^+$ and $\upmu^-$ beam polarisations do not have equal magnitude. Including the sin$\phi$ term in the fit
leads to
a very small change
in $\frac{\textrm{d}\sigma_\mr{LT}}    {\textrm{d}t}$,
which is compatible with zero within statistical uncertainties, while there is no impact on the determination of  $\frac{\textrm{d}\sigma_\mr{T}}{\textrm{d}t}+\epsilon \frac{\textrm{d}\sigma_\mr{L}}{\textrm{d}t}$ and
$\frac{\textrm{d}\sigma_\mr{TT}}    {\textrm{d}t}$.

}
\begin{table}[h!]
\centering
\caption[systematics]{Summary of the
estimated relative systematic uncertainties on the measured $|t|$ and $\phi$-dependent cross sections and on the extracted cross-section contributions
$\frac{\textrm{d}\sigma_\mr{U}}    {\textrm{d}t} =\frac{\textrm{d}\sigma_\mr{T}}{\textrm{d}t}+\epsilon \frac{\textrm{d}\sigma_\mr{L}}{\textrm{d}t}$   and
$\frac{\textrm{d}\sigma_\mr{TT}}    {\textrm{d}t}$
 in the full kinematic range.
 The values are given as a percentage. Note that the uni-directional uncertainty
$\sigma_{\uparrow}$ ($\sigma_{\downarrow}$) has to be used with positive (negative) sign.
}
\label{tab:systematics}
\centering
\resizebox{14cm}{!}{%
\begin{tabular}{  c   r    r    r    r   r  r r r  }
\hline \\ [-8pt]
  source &  $\sigma^t_{\uparrow}$ & $\sigma^t_{\downarrow}$ & $\sigma^{\phi}_{\uparrow}$ & $\sigma^{\phi}_{\downarrow}$
  &$\sigma_{\mr{U}\uparrow}$&$\sigma_{\mr{U}\downarrow}$
  &$\sigma_{\mr{TT}\uparrow}$&$\sigma_{\mr{TT}\downarrow}$\\[2pt]
\toprule
 $\upmu^+$ flux  &    2    & 2 & 2 & 2 & 2 & 2 & 2 & 2 \\[1pt]
 $\upmu^-$ flux  &    2    & 2 & 2 & 2 & 2 & 2 & 2 & 2 \\[2pt]
 \hline\\[-6pt]
  acceptance &         4  & 4 & 4 & 4 & 4 & 4 & 4 & 4\\[1pt]
  {ECAL0 threshold} &     5 -- 7  & 1 & 4 -- 8 & 1 & 5 & 1 & 4 & 1\\[1pt]
  {ECAL1 threshold} &     1 -- 2  & 1 & 1 -- 3 & 1 & 1 & 1 & 1 & 1\\[2pt]
 \hline\\[-6pt]
  $\chi^2$ of kinematic fit &      3     & 5 & 2.0 -- 5.6 & 4.0 -- 8.8 & 3 & 5 & 3 & 4\\[1pt]
 \textsc{Lepto} background & 6 -- 10 & 6 -- 10 & 6 -- 16 & 6 -- 16 & 8.3 & 8.3 & 1 & 1\\[1pt]
 \textsc{Lepto} normalisation & 2 -- 3 & 2 -- 3 & 2 -- 5 & 2 -- 5 & 2.6 & 2.6 & 2 & 2 \\[1pt]
 $\upomega$ background &  0 & 1.5 -- 2.7 & 0 & 1.4 -- 5.7 & 0 & 2.4 & 0 & 2.4 \\[2pt]
  \hline\\[-6pt]
   radiative corrections & 6 & 3 & 6.3 & 3.6 & 6 & 3 & 2 & 2\\[2pt]
 \hline\\[-6pt]
 $\sum$ & 12 -- 16 & 10.1 -- 13.1  & 11.6 -- 22.4  & 9.6 -- 20.1 & 13.3 & 11.7 &  7.7 & 7.1 \\[1pt]
\bottomrule\\[-6pt]
\end{tabular}}
\end{table}
\section{Results for the spin-independent cross section }
\label{sec:new-results}

\begin{figure}[h!]
\centering
\includegraphics[width=0.49\textwidth]{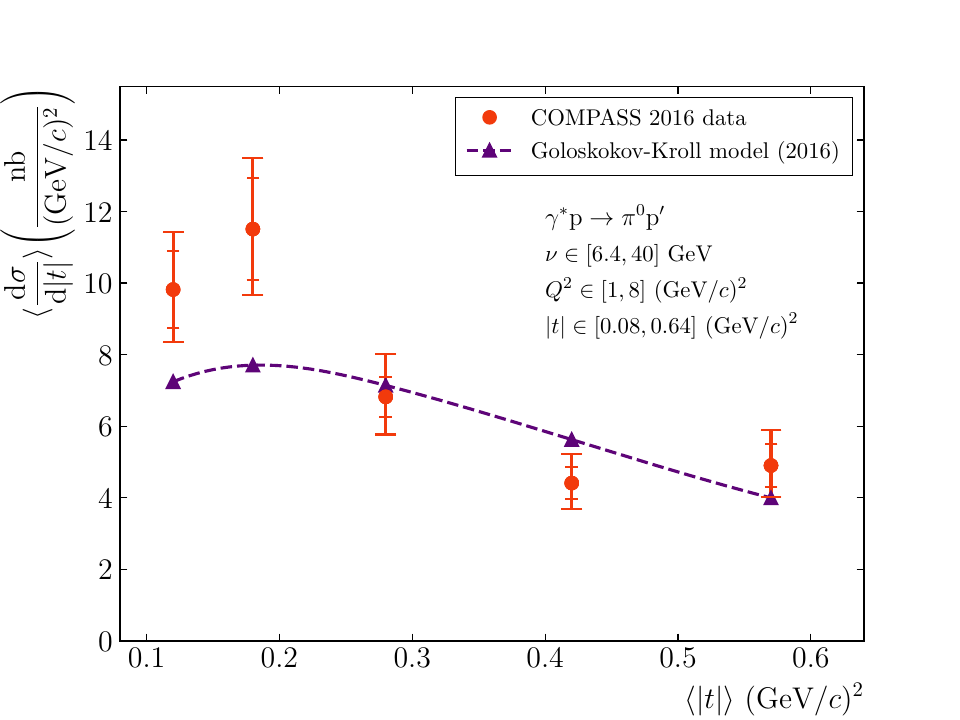}
\includegraphics[width=0.49\textwidth]{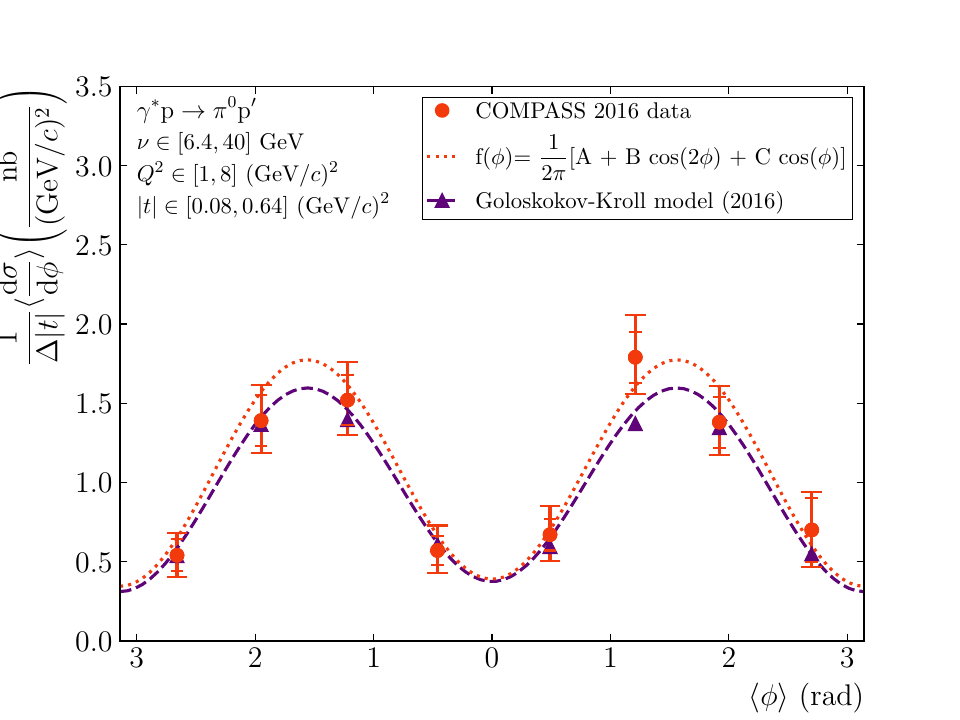}
\caption{\label{fig:x_sec_phi_t_2016}
    Left: spin-independent virtual-photon cross section integrated over the full $2\pi$-range in $\phi$, presented as a function of $|t|$. Right: spin-independent virtual-photon cross section averaged over the measured $|t|$-range, presented as a function of $\phi$.
   The inner error bars indicate the statistical uncertainty, the outer error bars the quadratic sum of statistical and systematic uncertainties. The triangles connected by the dashed line represent the Goloskokov--Kroll predictions~\cite{Goloskokov:2009ia, Goloskokov:2011rd, GKprivate, Matthias-Gorzellik} for the experimental bins and the  dotted line is a fit of the $\phi$ distribution using  Eq.~(\ref{eq:xsection_unpol}).
   }
\end{figure}

The dependences of the measured cross sections on $|t|$ and on $\phi$ are shown in  Fig.~\ref{fig:x_sec_phi_t_2016}, with the numerical values given in Table~\ref{tab:2016}. In order to obtain the $|t|$-dependence, the cross section is integrated
over $\phi$, while for the $\phi$-dependence it is averaged over $|t|$.
The kinematic domain of the measurement
is given by
0.08~\GeVcsq $< \, |t| \, <$  0.64~\GeVcsq,
1~\GeVcsq $< \, Q^2 \, <$ 8~\GeVcsq,
6.4~GeV $<\,\nu\,<$ 40~GeV,
while the average kinematics are:
$\langle |t| \rangle =$~0.29~\GeVcsq,
$\langle Q^2 \rangle =$~2.27~\GeVcsq; $\langle \nu \rangle =$~10.16~GeV, $\langle \xBj \rangle =$~0.134, $\langle W \rangle =$~4.1~\GeVcc, $\langle y \rangle =$~0.064 and $\langle \epsilon \rangle =$~0.997.

\begin{table}[h!]
\caption{\label{tab:2016}
Numerical values of the cross sections shown in Fig.~\ref{fig:x_sec_phi_t_2016} with the mean values of $|t|$ and $\phi$ in each bin.}
\centering
\resizebox{\textwidth}{!}{%
\begin{tabular}{c c c c l c c}
\hline
 &  & \hspace{1cm} & & & & \\[-4pt]
{\footnotesize $|t|$-range} & $\langle |t|\rangle$ $\bigl[\text{\GeVcsq}\bigr]$ & $\Big\langle \frac{\di \sigma}{\di |t|} \Big\rangle~\left[\frac{\text{nb}}{\text{\GeVcsq}} \right]$ & \hspace{0.5cm} & \hspace{0.5cm}{\footnotesize $\phi$-range } & $\langle \phi \rangle$ [rad] & $\frac{1}{\Delta |t| }\Big\langle \frac{\di \sigma}{\di \phi} \Big\rangle ~\left[\frac{\text{nb}}{\text{\GeVcsq}} \right]$  \\[8pt]
\hline\\[-6pt]
0.08 -- 0.15 & 0.12 & \fz\xsecvil{9.82}{1.08}{1.00}{1.18} & & \phibi    & $-$2.66 & \xsecvil{0.54}{0.10}{0.09}{0.10}  \\[8pt]
0.15 -- 0.22 & 0.18 & \xsecvil{11.51}{1.42}{1.16}{1.38}   & & \phibii   & $-$1.95 & \xsecvil{1.39}{0.16}{0.13}{0.16}  \\[8pt]
0.22 -- 0.36 & 0.28 & \fz\xsecvil{6.82}{0.56}{0.89}{1.06} & & \phibiii  & $-$1.22 & \xsecvil{1.52}{0.16}{0.15}{0.18}  \\[8pt]
0.36 -- 0.50 & 0.42 & \fz\xsecvil{4.41}{0.45}{0.58}{0.68} & & \phibiv   & $-$0.46 & \xsecvil{0.57}{0.09}{0.11}{0.13}  \\[8pt]
0.50 -- 0.64 & 0.57 & \fz\xsecvil{4.90}{0.60}{0.63}{0.78} & & \phibv    & \fm 0.49 & \xsecvil{0.67}{0.10}{0.13}{0.15}  \\[8pt]
             &      &                                     & & \phibvi   & \fm 1.21 & \xsecvil{1.79}{0.16}{0.17}{0.21}  \\[8pt]
0.08 -- 0.64 & 0.29 & \fz\xsecvil{6.72}{0.32}{0.79}{0.89} & & \phibvii  & \fm 1.92 & \xsecvil{1.38}{0.16}{0.13}{0.16}  \\[8pt]
             &      &                                     & & \phibviii & \fm 2.70 & \xsecvil{0.70}{0.20}{0.12}{0.13}   \\[8pt]
\hline
\end{tabular}}
\end{table}

The cross section $\langle \frac{\di \sigma}{\di |t|} \rangle$ shown in the left panel of Fig.~\ref{fig:x_sec_phi_t_2016}  decreases with increasing $|t|$ for values of   $|t|$ larger than 0.2~\GeVcc, while at smaller $|t|$  the trend seems to be the opposite.
The measured cross section is reasonably well described by the predictions of the Goloskokov--Kroll (GK) model from 2016~\cite{Goloskokov:2009ia, Goloskokov:2011rd, GKprivate}. Cross sections are calculated within the handbag approach, which is based on factorization in hard parton subprocesses and GPDs. The latter are constructed from double distributions with parameters given in~\cite{GKprivate, Matthias-Gorzellik}.
The result of the GK model is obtained by integrating over the same kinematic range as for the data.


The  cross section $\frac{1}{\Delta |t|}    \langle \frac{\di^2 \sigma}{ \di \phi} \rangle$ averaged over the full $|t|$-range
is shown in the right panel of Fig.~\ref{fig:x_sec_phi_t_2016}.
In order to extract the different contributions to the spin-independent cross section, a binned likelihood fit
is applied to the data according to Eq.~(\ref{eq:xsection_unpol}) using the integral of the fit function in each bin.
The average value of the virtual-photon polarisation parameter is used in the fit. The results are presented in Table~\ref{tab:fit_par_2016}.


\begin{table}[h!]
\caption{\label{tab:fit_par_2016}
The contributions in $\frac{\text{nb}}{\text{\GeVcsq}}$ to the spin-independent cross section.
}
\centering
\resizebox{14cm}{!}{
\begin{tabular}{c c c c c}
\hline \\[-6pt]
$\Big\langle \frac{\di \sigma_{\rm{T}}}{\di |t|} + \epsilon \frac{\di \sigma_{\rm{L}}}{\di |t|} \Big\rangle$ & \hspace{0.5cm}
& $ \Big\langle \frac{\di \sigma_{\rm{TT}}}{\di |t|} \Big\rangle $ &\hspace{0.5cm}
& $ \Big\langle \frac{\di \sigma_{\rm{LT}}}{\di |t|} \Big\rangle $  \\ [6pt]
\hline
& & \\
\xsecvil{6.7}{0.3}{0.8}{0.9} && \xsecvil{-4.4}{0.5}{0.3}{0.3} & & \xsecvil{0.1}{0.2}{0.1}{0.1} \\ [6pt]
\hline
\end{tabular}}
\end{table}

We observe a longitudinal--transverse interference contribution compatible with zero within statistical uncertainties, while the
transverse--transverse interference contribution is large and negative, and of the same order of magnitude as the sum of transverse and longitudinal contributions.
The exclusive $\uppi^0$ production cross section  depends on the GPDs $\widetilde{H}$, $\widetilde{E}$,
$H_\mr{T}$ and $\overline{E}_\mr{T}$ (see~\cref{equ::relation_L,equ::relation_T,equ::relation_TT}). As explained in the introduction,
 the contribution of the chiral-odd GPD $\overline{E}_\mr{T}$ is dominant due to the relative sign of $u$ and $d$ quark contributions  for the $\uppi^0$ production, in contrast to the contributions to the other GPDs.  Our observation of a large contribution from $\sigmaTT$ and a slight dip in the differential cross section
 $\langle \frac{\di \sigma}{\di |t|} \rangle$
 as $|t|$ decreases to zero supports this expectation, which is also described by the GK model~\cite{Goloskokov:2009ia, Goloskokov:2011rd, GKprivate, Matthias-Gorzellik}.

\section{Comparison to previously published results}
\label{sec:old-results}

\begin{figure}[h!]
\centering
\includegraphics[width=0.49\textwidth]{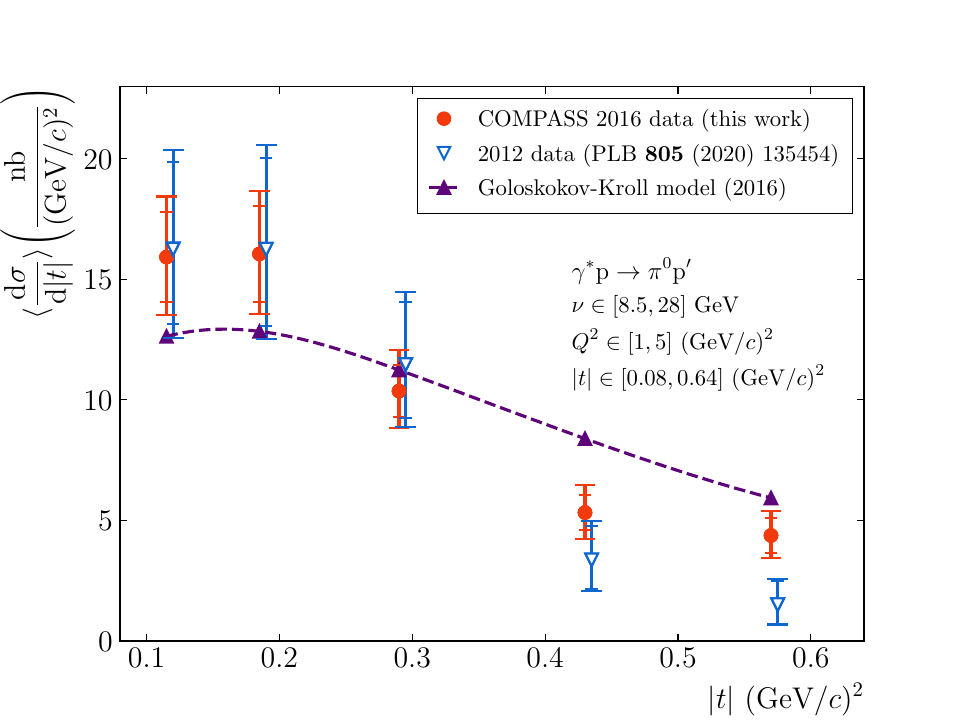}
\includegraphics[width=0.49\textwidth]{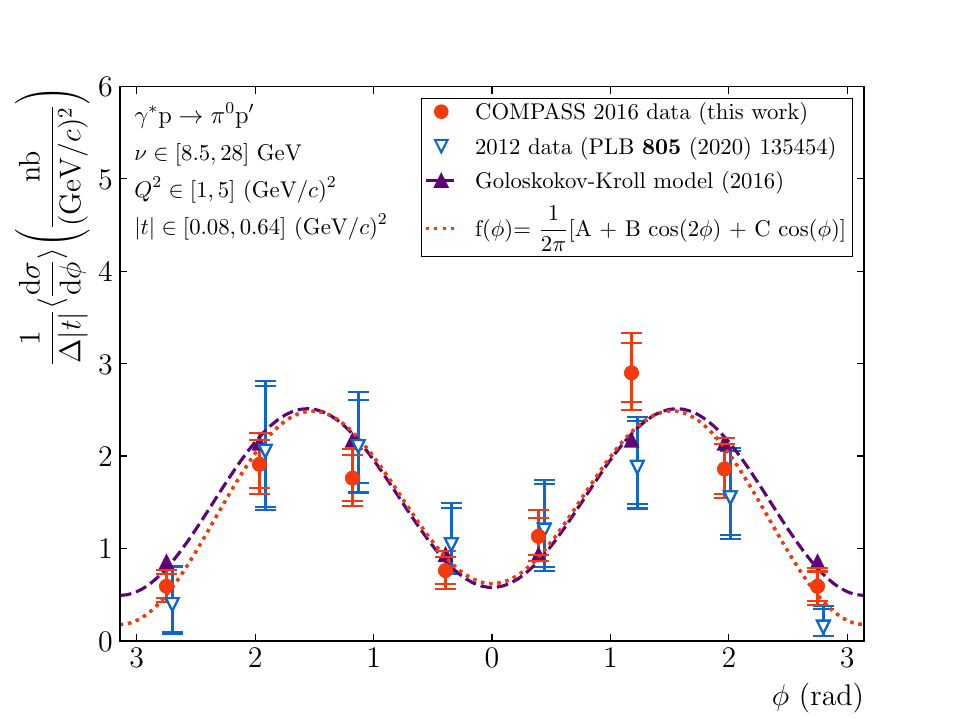}
  \caption{\label{fig:x_sec_phi_t_2012}
   Left: spin-independent virtual-photon cross section integrated over the full $2\pi$-range in $\phi$, presented as a function of $|t|$. Right: spin-independent virtual-photon cross section averaged over the measured $|t|$-range, presented as a function of $\phi$.
The cross sections obtained in the present measurement using the 2016 data (in red) are compared to the previous ones (in blue). The inner error bars indicate the statistical uncertainty, the outer error bars the quadratic sum of statistical and systematic uncertainties. The 2016 data points are shown at the corresponding bin centre. For better visibility the 2012 data points are shifted horizontally with respect to the present results. The triangles connected by the dashed line represent the Goloskokov--Kroll predictions~\cite{Goloskokov:2009ia, Goloskokov:2011rd, GKprivate, Matthias-Gorzellik} for the experimental bins and the dotted line is a fit to the $\phi$ distribution of the 2016 data using  Eq.~(\ref{eq:xsection_unpol}).
   }
\end{figure}

In order to check the compatibility with the results obtained using the COMPASS 2012 data~\cite{COMPASS:2019fea}, the analysis is also performed in the previously accessible kinematic domain:
0.08~\GeVcsq $< \,|t|\,<$ 0.64~\GeVcsq,
1~\GeVcsq $< \, Q^2 \, <$ 5~\GeVcsq,
8.5~GeV $<\,\nu \,<$ 28~GeV and the average kinematics for this comparison are
$\langle |t| \rangle =$~0.28~\GeVcc,
$\langle Q^2 \rangle =$~2.16~\GeVcc, $\langle \nu \rangle =$ 12.34~GeV, $\langle \xBj \rangle =$~0.103, $\langle W \rangle =$~4.61~\GeVcc, $\langle y \rangle =$~0.078, $\langle \epsilon \rangle =$~0.996.
The differential cross sections of exclusive $\uppi^{0}$ production for the two sets of data are presented as a~function of $|t|$ and $\phi$ in
 Fig.~\ref{fig:x_sec_phi_t_2012} and the numerical values using the 2016 data set are given in Table~\ref{tab:2012}.

\begin{table}[h!]
\caption{\label{tab:2012}
Numerical values of the  cross sections shown in Fig.~\ref{fig:x_sec_phi_t_2012} with the mean values of $|t|$ and $\phi$ in each bin.}
\centering
\resizebox{\textwidth}{!}{%
\begin{tabular}{c c c c l c c}
\hline
 &  & \hspace{1cm} & & & & \\[-4pt]
{\footnotesize $|t|$-range} & $\langle |t|\rangle$ $\bigl[\text{\GeVcsq}\bigr]$ & $\Big\langle \frac{\di \sigma}{\di |t|} \Big\rangle~\left[\frac{\text{nb}}{\text{\GeVcsq}} \right]$ & \hspace{0.5cm} & \hspace{0.5cm}{\footnotesize $\phi$-range } & $\langle \phi \rangle$ [rad] & $\frac{1}{\Delta |t| }\Big\langle \frac{\di \sigma}{\di \phi} \Big\rangle ~\left[\frac{\text{nb}}{\text{\GeVcsq}} \right]$  \\[8pt]
\hline\\[-6pt]
0.08 -- 0.15 & 0.12 & \xsecvil{15.92}{1.86}{1.51}{1.69}   & & \phibi    & $-$2.64    & \xsecvil{0.59}{0.13}{0.11}{0.12} \\[8pt]
0.15 -- 0.22 & 0.18 & \xsecvil{16.05}{1.98}{1.49}{1.70}   & & \phibii   & $-$1.97    & \xsecvil{1.91}{0.26}{0.19}{0.21} \\[8pt]
0.22 -- 0.36 & 0.28 & \xsecvil{10.36}{1.07}{1.12}{1.34}   & & \phibiii  & $-$1.23    & \xsecvil{1.76}{0.25}{0.17}{0.19} \\[8pt]
0.36 -- 0.50 & 0.42 & \xsecvil{\fm5.32}{0.73}{0.82}{0.90} & & \phibiv   & $-$0.40    & \xsecvil{0.76}{0.15}{0.13}{0.15} \\[8pt]
0.50 -- 0.64 & 0.57 & \xsecvil{\fm4.37}{0.74}{0.59}{0.69} & & \phibv    & \fm 0.49 & \xsecvil{1.13}{0.20}{0.17}{0.20} \\[8pt]
             &      &                                     & & \phibvi   & \fm 1.20 & \xsecvil{2.90}{0.32}{0.24}{0.29} \\[8pt]
0.08 -- 0.64 & 0.28 & \xsecvil{\fm9.04}{0.50}{0.99}{1.08} & & \phibvii  & \fm 1.90 & \xsecvil{1.86}{0.27}{0.17}{0.19} \\[8pt]
             &      &                                     & & \phibviii & \fm 2.67 & \xsecvil{0.59}{0.16}{0.12}{0.12} \\[8pt]
\hline
\end{tabular}}
\end{table}

Our results are compared to the prediction of
the GK model
\cite{Goloskokov:2009ia, Goloskokov:2011rd, GKprivate, Matthias-Gorzellik}.
The measured differential cross sections
 $\langle \frac{\di \sigma}{\di |t|} \rangle$ are found to be compatible. However, a milder decrease is observed in the present data.
  Hence the new result is slightly closer to the predictions.
 The $\phi$-distributions of the cross sections averaged over the $|t|$-range are in good agreement for the two data sets and also with the GK prediction. In conclusion, the measured cross sections and the extracted contributions from the fit using the present data (see Table~\ref{tab:fit_par_2012}) are statistically compatible with the results from Ref.~\cite{COMPASS:2019fea}.


\begin{table}[h!]
\caption{\label{tab:fit_par_2012}
The contributions in $\frac{\text{nb}}{\text{\GeVcsq}}$ to the spin-independent cross section in the  kinematic domain
of Ref.~\cite{COMPASS:2019fea}.
}
\centering
\resizebox{\textwidth}{!}{%
\begin{tabular}{r c c c c c}
\hline \\[-6pt]
& $\Big\langle \frac{\di \sigma_{\rm{T}}}{\di |t|} + \epsilon \frac{\di \sigma_{\rm{L}}}{\di |t|} \Big\rangle$ & \hspace{0.4cm}
& $ \Big\langle \frac{\di \sigma_{\rm{TT}}}{\di |t|} \Big\rangle $ & \hspace{0.4cm}
& $ \Big\langle \frac{\di \sigma_{\rm{LT}}}{\di |t|} \Big\rangle $  \\ [6pt]
\hline
& & & & & \\ [-2pt]
2016 data \hspace{0.5cm} & \xsecvil{9.0}{0.5}{1.0}{1.1} &  &\xsecvil{-6.6}{0.8}{0.5}{0.5} &  &\xsecvil{0.7}{0.3}{0.4}{0.4} \\[2pt]
2012 data \hspace{0.5cm} & \xsecvil{8.1}{0.9}{1.0}{1.1} &  &\xsecvil{-6.0}{1.3}{0.7}{0.7} &  &\xsecvil{1.4}{0.5}{0.2}{0.3}\\[6pt]
\hline \\
\end{tabular}}
\end{table}

\section{Study of {the \texorpdfstring{$\phi$}{phi}-dependent cross section} in different \texorpdfstring{$|t|$}{|t|}-ranges}

\begin{figure}[h!]
\centering
\includegraphics[width=0.55\textwidth]{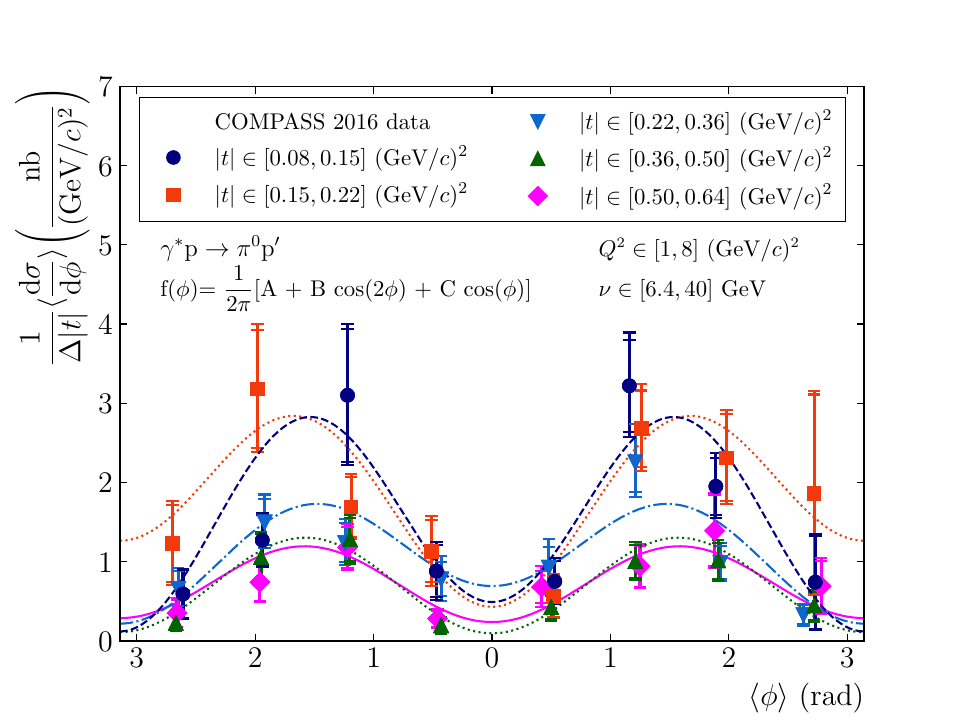 }
\caption{
\label{fig:xsection_phi_distr_5tbins}
 Spin-independent virtual-photon cross section in five $|t|$-ranges presented as a function of $\phi$.
The inner error bars indicate the statistical uncertainty, the outer error bars the quadratic sum of statistical and systematic uncertainties. The five curves are fits of the $\phi$ distributions using  Eq.~(\ref{eq:xsection_unpol}).
 }
\end{figure}

The $\phi$-modulation of the spin-independent cross section  is studied in five $|t|$-ranges using the full
$\nu$ and $Q^2$-ranges. The bin limits and the average kinematics are reported
in Table~\ref{tab:mean_values_tbins}.
The corresponding five differential cross sections of exclusive $\uppi^{0}$ production  are presented as a function of  $\phi$ in
 Fig.~\ref{fig:xsection_phi_distr_5tbins} and the numerical values  are given in Table~\ref{tab:app_t_1}.
\begin{table}[h!]
\caption{\label{tab:mean_values_tbins} Average values of the kinematic variables for the 5 $|t|$-bins using the full $\nu$ and $Q^2$-ranges.}
\centering
\resizebox{\textwidth}{!}{%
\begin{tabular}{ c c c c c c c c}
\hline \\[-10pt]
   $|t|$-range
 & $\langle Q^2 \rangle $ $\bigl[\text{\GeVcsq}\bigr]$
 & $\langle \nu \rangle$ [GeV]
 & $\langle |t|\rangle$ $\bigl[\text{\GeVcsq}\bigr]$
 & $\langle W \rangle$ [\GeVcc]
 & $\langle \xBj \rangle$
 & $\langle y \rangle$
 & $\langle \epsilon \rangle$ \\ [2pt]
    \hline
    0.08 -- 0.15 & 1.93 &   11.76 & 0.12 & 4.43  & 0.104 & 0.074  & 0.996 \\
    0.15 -- 0.22 & 2.11 &   10.32 & 0.18 & 4.16 & 0.123 & 0.065 & 0.997 \\
    0.22 -- 0.36 & 2.33 & \fz9.86 & 0.28 & 4.04  & 0.140 & 0.062 & 0.997 \\
    0.36 -- 0.50 & 2.41 & \fz9.29 & 0.42 & 3.92 & 0.150  & 0.059 &  0.998 \\
    0.50 -- 0.64 & 2.65 & \fz9.35 & 0.57 & 3.89 & 0.165 & 0.059 & 0.998  \\
       \hline \bigstrut
& $Q^2 \in [1, 8]$ & $\nu \in [6.4, 40]$  & & & & & \\
\hline
\end{tabular}}
\end{table}
%

\begin{figure}[h!]
\centering
\includegraphics[width=0.55\textwidth]{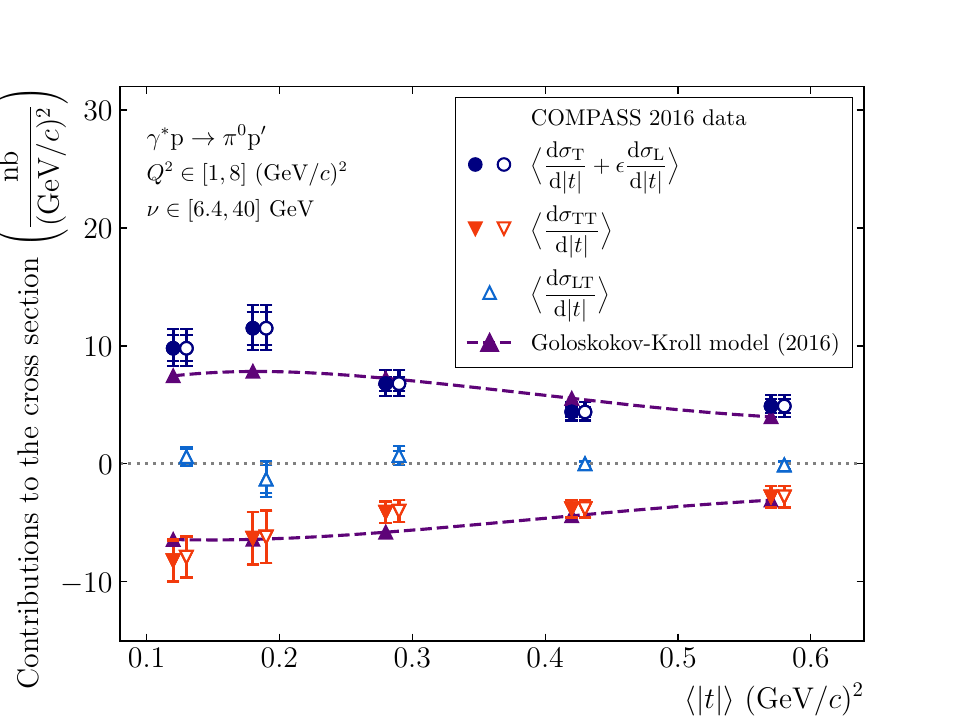 }
\caption{
\label{fig:extracted_xsection_5tbins}
    Extracted contributions to the cross section,
    $\frac{\textrm{d}\sigma_\mr{T}}{\textrm{d}t}+\epsilon \frac{\textrm{d}\sigma_\mr{L}}
    {\textrm{d}t}$,
    $\frac{\textrm{d}\sigma_\mr{TT}}
    {\textrm{d}t}$ and
    $\frac{\textrm{d}\sigma_\mr{LT}}
    {\textrm{d}t}$,
    as a function of $|t|$. Open points correspond to the fit of all three contributions, solid points correspond to the fit of two contributions with the assumption
    $\frac{\textrm{d}\sigma_\mr{LT}}
    {\textrm{d}t}=0$.
    The triangles connected by the dashed line represent the Goloskokov--Kroll predictions~\cite{Goloskokov:2009ia, Goloskokov:2011rd, GKprivate, Matthias-Gorzellik} for the experimental bins.
 }
\end{figure}

Fits of the $\phi$ distributions in the five $|t|$-ranges are applied using   Eq.~(\ref{eq:xsection_unpol}). The extracted contributions to the cross section, $\frac{\textrm{d}\sigma_\mr{T}}{\textrm{d}t}+\epsilon \frac{\textrm{d}\sigma_\mr{L}}
    {\textrm{d}t}$,
    $\frac{\textrm{d}\sigma_\mr{TT}}
    {\textrm{d}t}$ and
    $\frac{\textrm{d}\sigma_\mr{LT}}
    {\textrm{d}t}$,
are presented as a function of $|t|$ in Fig.\ref{fig:extracted_xsection_5tbins}. Open points correspond to the fit of the three contributions.
The contribution $\frac{\textrm{d}\sigma_\mr{LT}}{\textrm{d}t}$ is found compatible with zero within the statistical uncertainties.
A fit of only the two first contributions assuming $\frac{\textrm{d}\sigma_\mr{LT}}{\textrm{d}t} = 0$ is shown as solid points.
The results are given in
Table~\ref{tab:extracted_xsection_5tbins}.
We observe large and opposite contributions of $\frac{\textrm{d}\sigma_\mr{T}}{\textrm{d}t}+\epsilon \frac{\textrm{d}\sigma_\mr{L}}
    {\textrm{d}t}$ and
    $\frac{\textrm{d}\sigma_\mr{TT}}{\textrm{d}t}$.
Their absolute values  decrease with $|t|$ for $|t| >$~0.2~\GeVcc and do not exclude a small decrease towards smaller $|t|$. This confirms the dominance of the chiral-odd GPD $\overline{E}_T$ compared to the other involved GPDs
(see~\cref{equ::relation_L,equ::relation_T,equ::relation_TT}).


\begin{table}[htb]
\caption{\label{tab:extracted_xsection_5tbins}
The extracted contributions to the cross section in  ${\text{nb}}/{\text{\GeVcsq}}$ for the five $|t|$-ranges.}
\centering
\resizebox{14cm}{!}{%
\begin{tabular}{l c c c c }
\hline \\[-6pt]
\multicolumn{2}{l}{$|t|$-range $\bigl[\text{\GeVcsq}\bigr]$}
& $\Big\langle \frac{\di \sigma_{\rm{T}}}{\di |t|} + \epsilon \frac{\di \sigma_{\rm{L}}}{\di |t|} \Big\rangle$
& \hspace{0.5cm}
& $ \Big\langle \frac{\di \sigma_{\rm{TT}}}{\di |t|} \Big\rangle $   \\ [6pt]
\hline
 & & & &  \\ [-2pt]
0.08 -- 0.15  & & \fz\xsecvil{9.8}{1.1}{1.0}{1.2} & &\xsecvil{-8.2}{1.7}{0.6}{0.6}  \\
0.15 -- 0.22  & & \xsecvil{11.5}{1.4}{1.2}{1.4}   & &\xsecvil{-6.3}{2.2}{0.5}{0.5}  \\
0.22 -- 0.36  & & \fz\xsecvil{6.8}{0.6}{0.9}{1.0} & &\xsecvil{-4.1}{0.9}{0.3}{0.3}  \\
0.36 -- 0.50  & & \fz\xsecvil{4.4}{0.4}{0.6}{0.7} & &\xsecvil{-3.8}{0.7}{0.3}{0.3}  \\
0.50 -- 0.64  & & \fz\xsecvil{4.9}{0.6}{0.7}{0.7} & &\xsecvil{-2.8}{0.9}{0.2}{0.2}  \\[6pt]
\hline \\[-6pt]
\end{tabular}
}
\end{table}

\section{Dependence of the cross section on photon virtuality \texorpdfstring{$Q^2$}{Q2} and photon energy \texorpdfstring{$\nu$}{nu}}
\label{sec:new-results_Q2}

\begin{figure}[h!]
\centering
\includegraphics[width=0.49\textwidth]{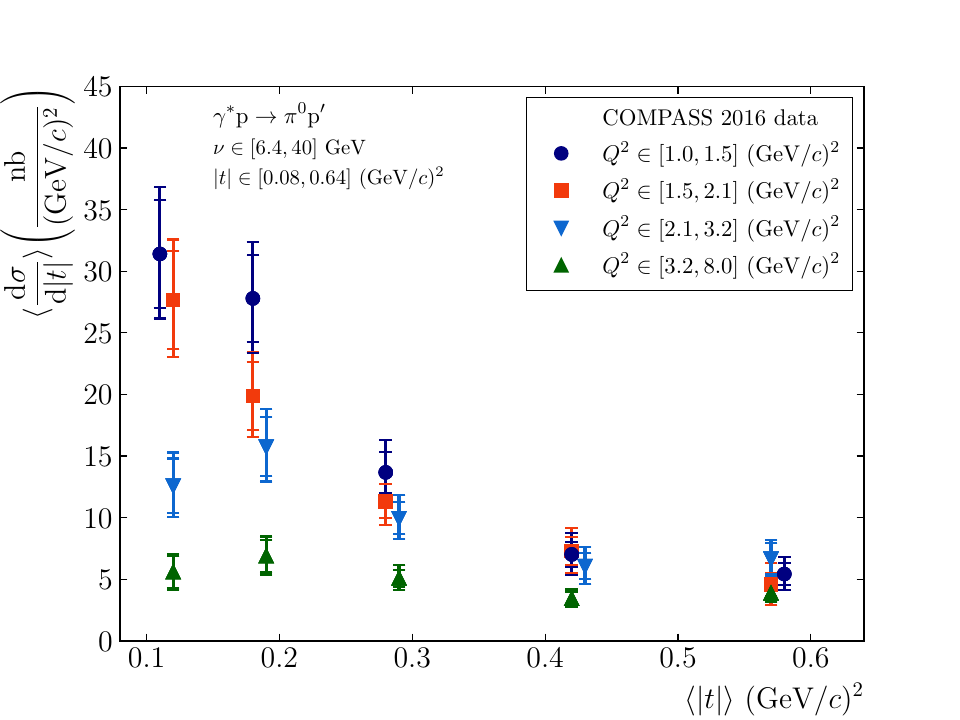}
\includegraphics[width=0.49\textwidth]{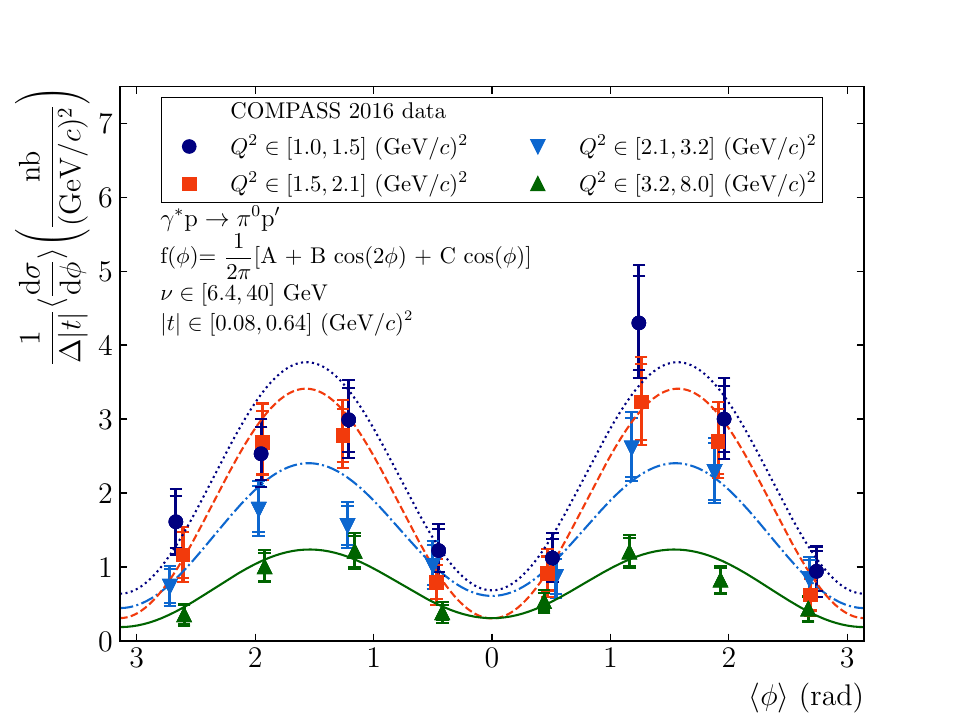}
  \caption{\label{Fig:Evolution_Q2}
       Left: spin-independent virtual-photon cross section integrated over the full $2\pi$-range in $\phi$, presented as a function of $|t|$.  Right: spin-independent virtual-photon cross section averaged over the measured $|t|$-range, presented as a function of $\phi$. Both figures show the results for the four $Q^2$-ranges.
       The inner error bars indicate the statistical uncertainty, the outer error bars the quadratic sum of statistical and systematic uncertainties. The four curves are fits of the $\phi$ distributions using  Eq.~(\ref{eq:xsection_unpol}).
   }
\end{figure}

\begin{figure}[h!]
\centering
\includegraphics[width=0.49\textwidth]{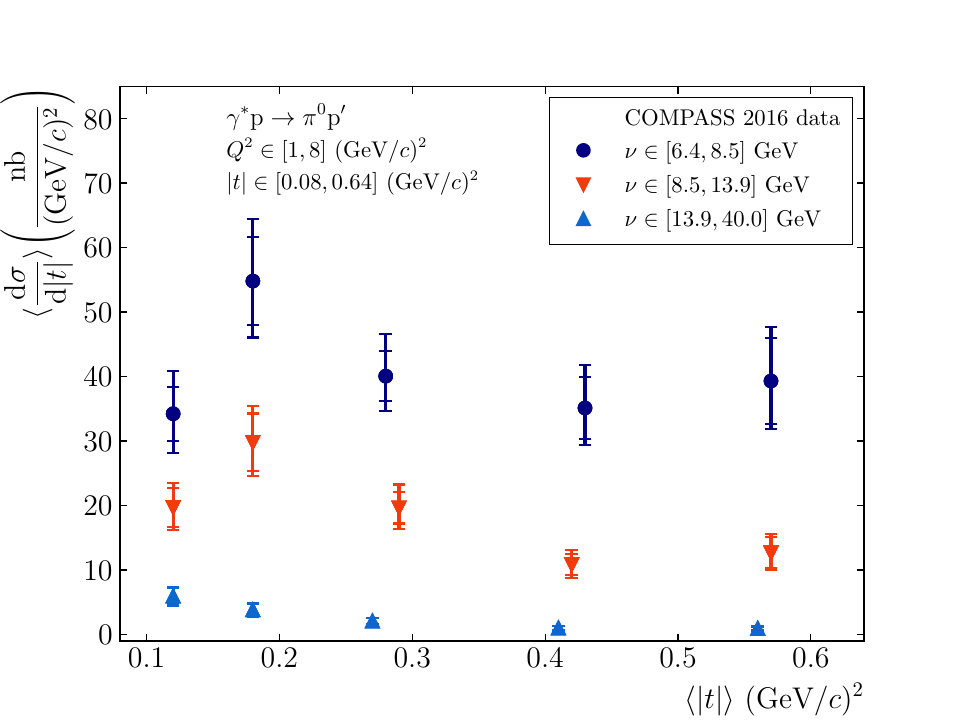}
\includegraphics[width=0.49\textwidth]{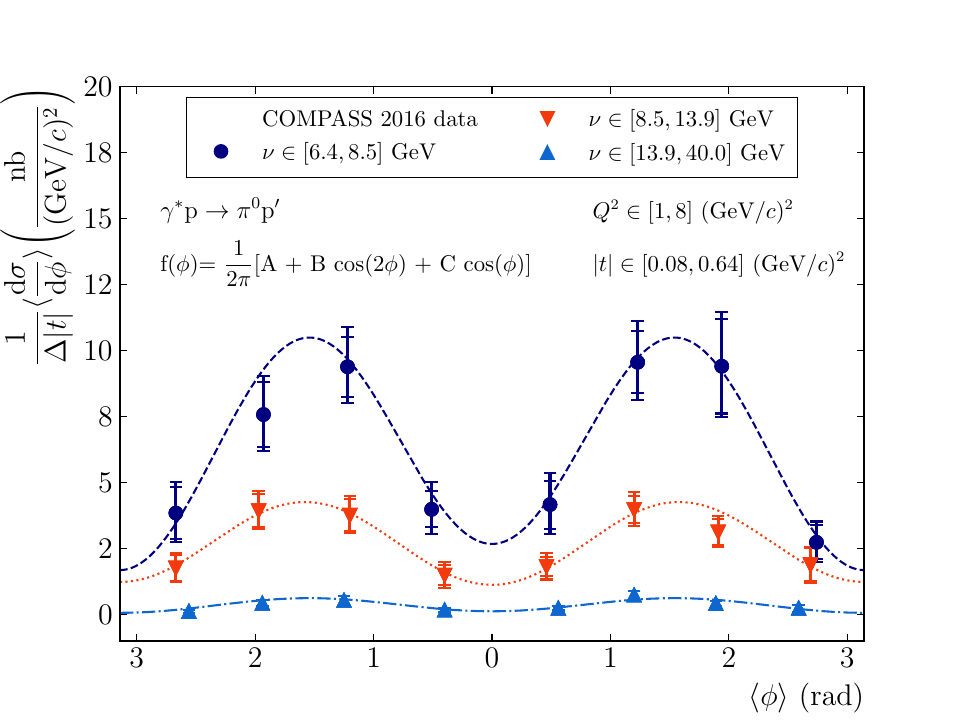}
  \caption{\label{Fig:Evolution_nu}
         Left: spin-independent virtual-photon cross section integrated over the full $2\pi$-range in $\phi$, presented as a function of $|t|$. Right: spin-independent virtual-photon cross section averaged over the measured $|t|$-range, presented as a function of $\phi$. Both figures show the results for the three $\nu$-ranges.
       The inner error bars indicate the statistical uncertainty, the outer error bars the quadratic sum of statistical and systematic uncertainties. The three curves are fits of the $\phi$ distributions using  Eq.~(\ref{eq:xsection_unpol}).
   }
\end{figure}

The spin-independent cross section  is also studied in four $Q^2$-ranges using the full
$\nu$ and $|t|$-ranges (see Fig.~\ref{Fig:Evolution_Q2} ), and in three $\nu$-ranges using the full
$Q^2$ and $|t|$-ranges (see Fig.~\ref{Fig:Evolution_nu} ). All numerical values  are given in \cref{tab:app_Q2_1,tab:app_Q2_2,tab:app_Q2_3,tab:app_Q2_4} and \cref{tab:app_nu_1,tab:app_nu_2,tab:app_nu_3}. The corresponding kinematic variables are reported in Tables~\ref{tab:mean_values_Q2bins} and \ref{tab:mean_values_nubins}.

\begin{table}[h!]
\caption{\label{tab:mean_values_Q2bins}
 Values of the kinematic variables for the four $Q^2$-ranges using the full $\nu$ and $|t|$-ranges. }
\centering
\resizebox{\textwidth}{!}{%
\begin{tabular}{ c c c c c c c c}
\hline \\[-10pt]
   $Q^2$-range
 & $\langle Q^2 \rangle $ $\bigl[\text{\GeVcsq}\bigr]$
 & $\langle \nu \rangle$ [GeV]
 & $\langle |t|\rangle$ $\bigl[\text{\GeVcsq}\bigr]$
 & $\langle W \rangle$ [\GeVcc]
 & $\langle \xBj \rangle$ & $\langle y \rangle$
 & $\langle \epsilon \rangle$ \\ [2pt]
    \hline
    1.0 -- 1.5 & 1.22 &   10.54 & 0.27 & 4.29  & 0.072 & 0.067 & 0.997 \\
    1.5 -- 2.1 & 1.77 & \fz9.81 & 0.27 & 4.09  & 0.109 & 0.062 & 0.997 \\
    2.1 -- 3.2 & 2.58 & \fz9.82 & 0.31 & 4.00  & 0.157 & 0.062 & 0.997 \\
    3.2 -- 8.0 & 4.33 &   10.39 & 0.33 & 3.90  & 0.247 & 0.065 & 0.997 \\
       \hline \bigstrut
& & $\nu \in [6.4, 40]$  &
$|t| \in [0.08, 0.64]$ & & & & \\
\hline
\end{tabular}}
\end{table}

 \begin{table}[h!]
\caption{\label{tab:mean_values_nubins} Values of the kinematic variables for the three $\nu$-bins
using the full $|t|$ and $Q^2$-ranges.}
\centering
\resizebox{\textwidth}{!}{%
\begin{tabular}{ l c c c c c c c}
\hline \\ [-10pt]
 $\nu$-range
 & $\langle Q^2 \rangle $ $\bigl[\text{\GeVcsq}\bigr]$ & $\langle \nu \rangle$ [GeV] & $\langle |t|\rangle$ $\bigl[\text{\GeVcsq}\bigr]$ & $\langle W \rangle$ [\GeVcc]  & $\langle \xBj \rangle$ & $\langle y \rangle$ & $\langle \epsilon \rangle$ \\ [2pt]
\hline
\fz6.4 -- \fz8.5 & 2.15 & \fz7.35 & 0.31 & 3.53  & 0.156 & 0.046 & 0.999 \\
\fz8.5 --   13.9 & 2.50 &   10.32 & 0.29 & 4.20  & 0.131 & 0.065 & 0.998 \\
          13.9 --   40.0 & 2.09 &   21.08 & 0.24 & 6.12  & 0.057 & 0.133 & 0.989 \\
\hline \bigstrut
       &  $Q^2 \in [1, 8]$ & &$|t| \in [0.08, 0.64]$ & & & & \\
\hline
\end{tabular}}
\end{table}

The differential cross sections studied as a function of $|t|$  (left part of Fig.~\ref{Fig:Evolution_Q2} and~\ref{Fig:Evolution_nu}) show a maximum around $|t| =$~0.2~\GeVcsq, as it was already mentioned.
However, no maximum  is observed
at small $Q^2$ or large $\nu$, which corresponds to $\xBj <$~0.1 (see Tables~\ref{tab:mean_values_Q2bins} and \ref{tab:mean_values_nubins}). Here, significant longitudinal contributions
are expected, which decrease with increasing $|t|$. This was already emphasised in Ref.~ \cite{Duplancic:2023xrt}.

\begin{figure}[h!]
\centering
\includegraphics[width=0.49\textwidth]{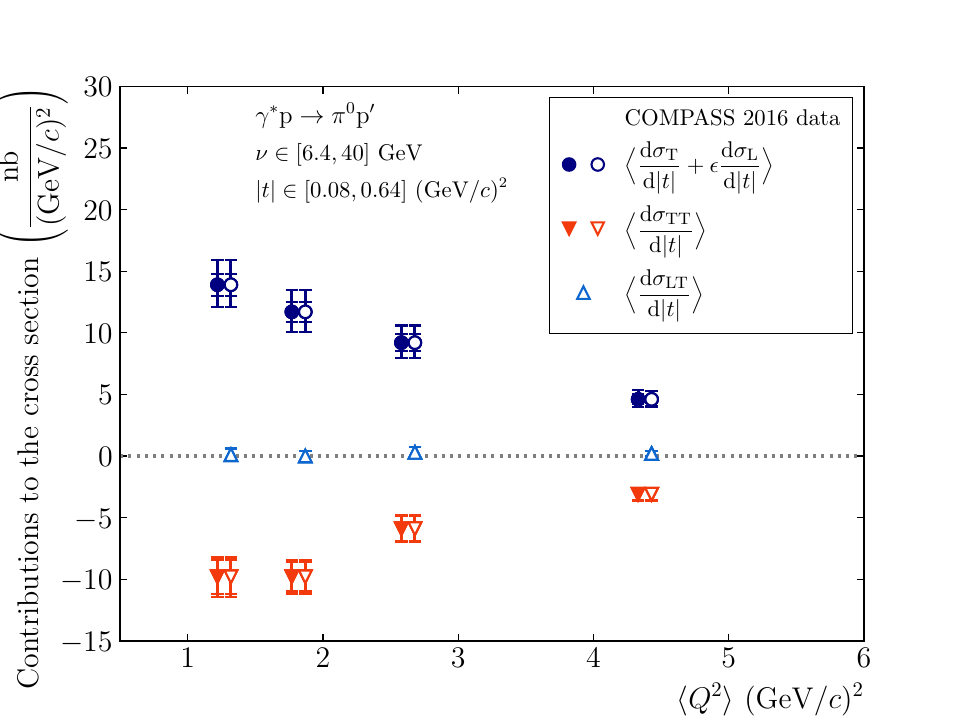 }
\includegraphics[width=0.49\textwidth]{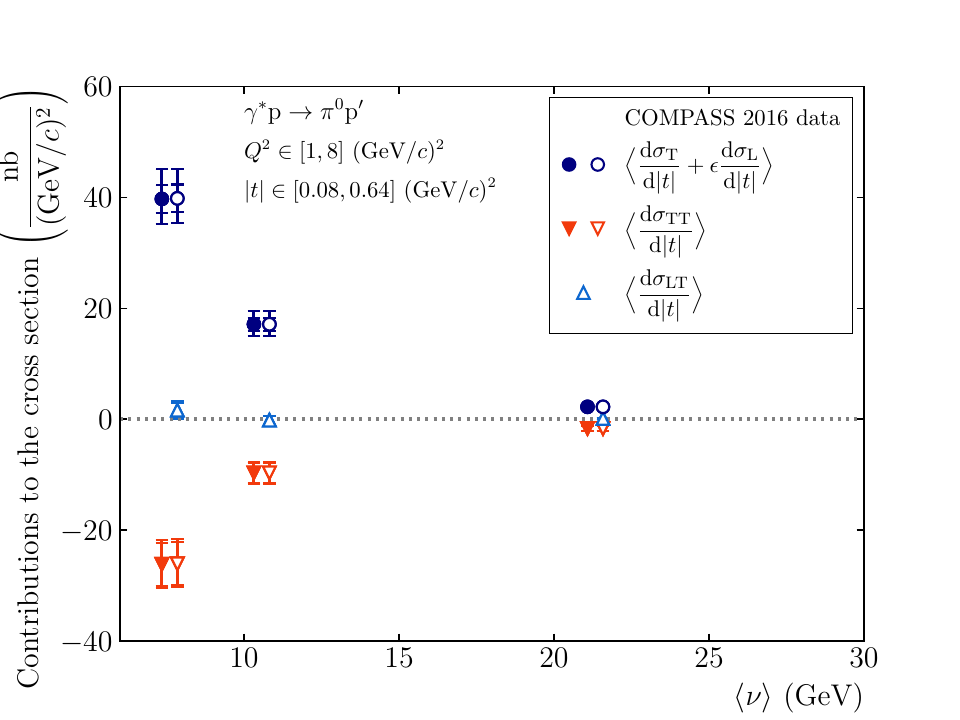 }
\caption{
\label{fig:extracted_xsection_4Q2bins_3nubins}
Extracted contributions to the cross section,
    $\frac{\textrm{d}\sigma_\mr{T}}{\textrm{d}t}+\epsilon \frac{\textrm{d}\sigma_\mr{L}}
    {\textrm{d}t}$,
    $\frac{\textrm{d}\sigma_\mr{TT}}
    {\textrm{d}t}$ and
    $\frac{\textrm{d}\sigma_\mr{LT}}
    {\textrm{d}t}$,
    as a function of $Q^2$ (left) and $\nu$ (right). Open points correspond to the fit of the three contributions, solid points correspond to the fit of two contributions with the assumption
    $\frac{\textrm{d}\sigma_\mr{LT}}
    {\textrm{d}t}=0$.
}
\end{figure}

The differential cross section is shown as a function of $\phi$  in the right parts of Fig.~\ref{Fig:Evolution_Q2} (in the four $Q^2$-ranges) and  of Fig.~\ref{Fig:Evolution_nu} (in the three $\nu$-ranges) together with fits using Eq.~(\ref{eq:xsection_unpol}).
The contributions  to the cross section,
    $\frac{\textrm{d}\sigma_\mr{T}}{\textrm{d}t}+\epsilon \frac{\textrm{d}\sigma_\mr{L}}
    {\textrm{d}t}$,
    $\frac{\textrm{d}\sigma_\mr{TT}}
    {\textrm{d}t}$ and
    $\frac{\textrm{d}\sigma_\mr{LT}}
    {\textrm{d}t}$,
obtained from the fits,
are presented as a function of $Q^2$ and $\nu$ in Fig.~\ref{fig:extracted_xsection_4Q2bins_3nubins}.
As in Fig.~{\ref{fig:extracted_xsection_5tbins}}, results with two or three contributions in the fit are presented.
Again
no differences are observed using the assumption
$\frac{\textrm{d}\sigma_\mr{LT}}
    {\textrm{d}t} = 0$.
The numerical values are given in Table~\ref{tab:extracted_xsection_4Q2bins_3nubins}.  A strong decrease of the absolute values of  $\frac{\textrm{d}\sigma_\mr{T}}{\textrm{d}t}+\epsilon \frac{\textrm{d}\sigma_\mr{L}}
    {\textrm{d}t}$ and
    $\frac{\textrm{d}\sigma_\mr{TT}}
    {\textrm{d}t}$
with increasing $\nu$ is observed, while the decrease is weaker with increasing $Q^2$.
These results provide valuable input for modelling GPDs and exclusive $\uppi^0$ production cross sections.
Furthermore, they should facilitate the study of next-to-leading order corrections and higher-twist contributions (as proposed in Ref.~\cite{Duplancic:2023xrt}).

\begin{table}[ht]
\caption{\label{tab:extracted_xsection_4Q2bins_3nubins}
The extracted contributions to the cross section in  ${\text{nb}}/{\text{\GeVcsq}}$ for the 4 $Q^2$-ranges and for the 3 $\nu$-ranges}
\centering
\resizebox{12cm}{!}{%
\begin{tabular}{l c c c c }
\hline \\[-6pt]
\multicolumn{2}{r}{$Q^2$-range $\bigl[\text{\GeVcsq}\bigr]$}
& $\Big\langle \frac{\di \sigma_{\rm{T}}}{\di |t|} + \epsilon \frac{\di \sigma_{\rm{L}}}{\di |t|} \Big\rangle$
& \hspace{0.5cm}
& $ \Big\langle \frac{\di \sigma_{\rm{TT}}}{\di |t|} \Big\rangle $   \\ [6pt]
\hline
 & & & &  \\ [-2pt]
1.0 -- 1.5  &&   \xsecvil{13.9}{0.9}{1.6}{1.8} && \xsecvil{-9.8}{1.2}{0.7}{0.7}  \\
1.5 -- 2.1  &&   \xsecvil{11.7}{0.8}{1.4}{1.6} && \xsecvil{-9.8}{1.2}{0.7}{0.7}  \\
2.1 -- 3.2  && \fz\xsecvil{9.2}{0.7}{1.0}{1.2} && \xsecvil{-5.9}{1.0}{0.4}{0.4}  \\
3.2 -- 8.0  && \fz\xsecvil{4.6}{0.4}{0.5}{0.6} && \xsecvil{-3.1}{0.5}{0.2}{0.2}  \\
[6pt]
\hline \\[-6pt]
 \multicolumn{2}{l}{$\nu$-range $[{\rm GeV}]$} & & &   \\[6pt]
 \hline
 & & & &  \\ [-2pt]
\fz6.4 -- \fz8.5 && \xsecvil{39.7}{2.5}{3.7}{4.7} && \xsecvil{-26.2}{3.9}{1.7}{2.0} \\
\fz8.5 --   13.9 && \xsecvil{17.1}{1.2}{1.7}{2.1} && \xsecvil{-9.7}{1.8}{0.7}{0.7}  \\
  13.9 --   40.0 && \fz\xsecvil{2.2}{0.2}{0.3}{0.3}  && \xsecvil{-1.7}{0.4}{0.2}{0.1}  \\
\hline
\end{tabular}}
\end{table}


\section{Summary and conclusion}
\label{sec:conlusion}

We have measured the $t$ and $\phi$-dependence of the spin-independent virtual-photon--proton cross section using hard exclusive $\uppi^0$ muoproduction with 
$\upmu^{+}$ and $\upmu^{-}$ beams in a wide kinematic range with the photon virtuality ranging from 1 to 8~\GeVcsq and the photon energy from 6.4 to 40 GeV. Fitting the azimuthal distribution in different $|t|$-ranges from 0.08 to 0.64~\GeVcsq shows a large negative transverse-transverse interference  contribution opposite to the sum of the transversely and longitudinally polarised virtual-photon contributions. This is clear evidence of an impact of the chiral-odd GPD $\overline{E}_{T}$.
The $|t|$-dependence at small values of $\xBj$ indicates that the longitudinally polarised virtual-photon contribution is not negligible. The measured dependences of these contributions on $Q^2$ and $\nu$ provide valuable input for new model calculations including higher twists and next-to-leading order corrections.


\section*{Acknowledgements}
We gratefully acknowledge the support of the CERN
management and staff and the skill and effort of the technicians of
our collaborating institutes. We would like to express our gratitude to Peter Kroll and Kornelia Passek-Kumericki for their continuous theoretical support.

\clearpage
\appendix
\section{Appendix}

\begin{table}[h!]
\caption{\label{tab:app_Q2_1}
Numerical values of the cross sections shown in Fig.~\ref{Fig:Evolution_Q2} for
1\,\text{\GeVcsq} $< Q^2 <$ 1.5\,\text{\GeVcsq}  with the mean values of $|t|$ and $\phi$ in each bin.}
\centering
\resizebox{\textwidth}{!}{%
\begin{tabular}{c c c c c c c}
\hline
 &  & \hspace{1cm} & & & & \\[-4pt]
{\footnotesize $|t|$-range} & $\langle |t|\rangle$ $\bigl[\text{\GeVcsq}\bigr]$ & $\Big\langle \frac{\di \sigma}{\di |t|} \Big\rangle~\left[\frac{\text{nb}}{\text{\GeVcsq}} \right]$ & \hspace{0.5cm} & {\footnotesize $\phi$-range } & $\langle \phi \rangle$ [rad] & $\frac{1}{\Delta |t| }\Big\langle \frac{\di \sigma}{\di \phi} \Big\rangle ~\left[\frac{\text{nb}}{\text{\GeVcsq}} \right]$  \\[8pt]
\hline\\[-6pt]
0.08 -- 0.15 & 0.11 & \xsecvil{31.40}{4.37}{2.89}{3.20} & & \phibi & $-2.67$ & \xsecvil{1.61}{0.35}{0.27}{0.28} \\[8pt]
0.15 -- 0.22 & 0.18 & \xsecvil{27.79}{3.52}{2.67}{2.97} & & \phibii & $-1.95$ & \xsecvil{2.53}{0.36}{0.27}{0.30} \\[8pt]
0.22 -- 0.36 & 0.28 & \xsecvil{13.67}{1.68}{1.72}{1.98} & & \phibiii & $-1.21$ & \xsecvil{2.99}{0.43}{0.29}{0.33} \\[8pt]
0.36 -- 0.50 & 0.42 & \fz\xsecvil{7.01}{1.04}{1.27}{1.35} & & \phibiv & $-0.45$ & \xsecvil{1.22}{0.29}{0.18}{0.21} \\[8pt]
0.50 -- 0.64 & 0.58 & \fz\xsecvil{5.42}{0.92}{0.93}{1.04} & & \phibv & $\fm0.51$ & \xsecvil{1.12}{0.26}{0.20}{0.22} \\[8pt]
             &      &                                     & & \phibvi & $\fm1.24$ & \xsecvil{4.30}{0.64}{0.37}{0.45} \\[8pt]
0.08 -- 0.64 & 0.27 & \xsecvil{13.91}{0.89}{1.59}{1.71}   & & \phibvii & $\fm1.96$ & \xsecvil{3.00}{0.45}{0.30}{0.33} \\[8pt]
             &      &                                     & & \phibviii & $\fm2.74$ & \xsecvil{0.94}{0.27}{0.21}{0.20} \\[8pt]
\hline
\end{tabular}}
\end{table}

\begin{table}[h!]
\caption{\label{tab:app_Q2_2}
Numerical values of the  cross sections shown in Fig.~\ref{Fig:Evolution_Q2} for
1.5\,\text{\GeVcsq}$<  Q^2 <$  2.1\,\text{\GeVcsq} with the mean values of $|t|$ and $\phi$ in each bin.}
\centering
\resizebox{\textwidth}{!}{%
\begin{tabular}{c c c c c c c}
\hline
 &  & \hspace{1cm} & & & & \\[-4pt]
{\footnotesize $|t|$-range} & $\langle |t|\rangle$ $\bigl[\text{\GeVcsq}\bigr]$ & $\Big\langle \frac{\di \sigma}{\di |t|} \Big\rangle~\left[\frac{\text{nb}}{\text{\GeVcsq}} \right]$ & \hspace{0.5cm} & {\footnotesize $\phi$-range } & $\langle \phi \rangle$ [rad] & $\frac{1}{\Delta |t| }\Big\langle \frac{\di \sigma}{\di \phi} \Big\rangle ~\left[\frac{\text{nb}}{\text{\GeVcsq}} \right]$  \\[8pt]
\hline\\[-6pt]
0.08 -- 0.15 & 0.12 & \xsecvil{27.65}{3.99}{2.27}{2.88} & & \phibi & $-2.61$ & \xsecvil{1.16}{0.31}{0.19}{0.21} \\[8pt]
0.15 -- 0.22 & 0.18 & \xsecvil{19.87}{2.77}{1.83}{2.25} & & \phibii & $-1.94$ & \xsecvil{2.68}{0.43}{0.26}{0.31} \\[8pt]
0.22 -- 0.36 & 0.28 & \xsecvil{11.33}{1.36}{1.38}{1.69} & & \phibiii & $-1.26$ & \xsecvil{2.78}{0.36}{0.26}{0.31} \\[8pt]
0.36 -- 0.50 & 0.42 & \fz\xsecvil{7.28}{1.16}{1.38}{1.51} & & \phibiv & $-0.47$ & \xsecvil{0.79}{0.23}{0.20}{0.21} \\[8pt]
0.50 -- 0.64 & 0.57 & \fz\xsecvil{4.55}{0.94}{1.39}{1.46} & & \phibv & $\fm0.47$ & \xsecvil{0.91}{0.23}{0.22}{0.24} \\[8pt]
             &      &                                     & & \phibvi & $\fm1.26$ & \xsecvil{3.23}{0.51}{0.27}{0.34} \\[8pt]
0.08 -- 0.64 & 0.27 & \xsecvil{11.67}{0.79}{1.37}{1.56}   & & \phibvii & $\fm1.91$ & \xsecvil{2.70}{0.44}{0.23}{0.29} \\[8pt]
             &      &                                     & & \phibviii & $\fm2.69$ & \xsecvil{0.62}{0.21}{0.20}{0.21} \\[8pt]
\hline
\end{tabular}}
\end{table}

\begin{table}[h!]
\caption{\label{tab:app_Q2_3}
Numerical values of the cross sections shown in Fig.~\ref{Fig:Evolution_Q2} for
2.1\,\text{\GeVcsq} $<  Q^2 <  3.2$\,\text{\GeVcsq} with the mean values of $|t|$ and $\phi$ in each bin.
}
\centering
\resizebox{\textwidth}{!}{%
\begin{tabular}{c c c c c c c}
\hline
 &  & \hspace{1cm} & & & & \\[-4pt]
{\footnotesize $|t|$-range} & $\langle |t|\rangle$ $\bigl[\text{\GeVcsq}\bigr]$ & $\Big\langle \frac{\di \sigma}{\di |t|} \Big\rangle~\left[\frac{\text{nb}}{\text{\GeVcsq}} \right]$ & \hspace{0.5cm} & {\footnotesize $\phi$-range } & $\langle \phi \rangle$ [rad] & $\frac{1}{\Delta |t| }\Big\langle \frac{\di \sigma}{\di \phi} \Big\rangle ~\left[\frac{\text{nb}}{\text{\GeVcsq}} \right]$  \\[8pt]
\hline\\[-6pt]
0.08 -- 0.15 & 0.12 & \xsecvil{12.59}{2.21}{1.26}{1.54} & & \phibi & $-2.72$ & \xsecvil{0.74}{0.23}{0.13}{0.15} \\[8pt]
0.15 -- 0.22 & 0.19 & \xsecvil{15.76}{2.42}{1.47}{1.83} & & \phibii & $-1.97$ & \xsecvil{1.78}{0.31}{0.18}{0.22} \\[8pt]
0.22 -- 0.36 & 0.29 & \fz\xsecvil{9.94}{1.30}{1.01}{1.34} & & \phibiii & $-1.22$ & \xsecvil{1.56}{0.26}{0.16}{0.19} \\[8pt]
0.36 -- 0.50 & 0.43 & \fz\xsecvil{6.06}{1.05}{1.05}{1.18} & & \phibiv & $-0.50$ & \xsecvil{1.02}{0.27}{0.16}{0.19} \\[8pt]
0.50 -- 0.64 & 0.57 & \fz\xsecvil{6.66}{1.25}{0.65}{0.92} & & \phibv & $\fm0.54$ & \xsecvil{0.87}{0.24}{0.16}{0.18} \\[8pt]
             &      &                                     & &  \phibvi & $\fm1.18$ & \xsecvil{2.61}{0.40}{0.21}{0.27} \\[8pt]
0.08 -- 0.64 & 0.31 & \fz\xsecvil{9.20}{0.66}{0.99}{1.18} & & \phibvii & $\fm1.88$ & \xsecvil{2.29}{0.38}{0.19}{0.25} \\[8pt]
             &      &                                     & &  \phibviii & $\fm2.68$ & \xsecvil{0.84}{0.25}{0.14}{0.16} \\[8pt]
\hline
\end{tabular}}
\end{table}

\begin{table}[p]
\caption{\label{tab:app_Q2_4}
Numerical values of the cross sections shown in Fig.~\ref{Fig:Evolution_Q2} for
3.2\,\text{\GeVcsq} $<  Q^2 <$ 8.0\,\text{\GeVcsq} with the mean values of $|t|$ and $\phi$ in each bin.
}
\centering
\resizebox{\textwidth}{!}{%
\begin{tabular}{c c c c c c c}
\hline
 &  & \hspace{1cm} & & & & \\[-4pt]
{\footnotesize $|t|$-range} & $\langle |t|\rangle$ $\bigl[\text{\GeVcsq}\bigr]$ & $\Big\langle \frac{\di \sigma}{\di |t|} \Big\rangle~\left[\frac{\text{nb}}{\text{\GeVcsq}} \right]$ & \hspace{0.5cm} & {\footnotesize $\phi$-range } & $\langle \phi \rangle$ [rad] & $\frac{1}{\Delta |t| }\Big\langle \frac{\di \sigma}{\di \phi} \Big\rangle ~\left[\frac{\text{nb}}{\text{\GeVcsq}} \right]$  \\[8pt]
\hline\\[-6pt]
0.08 -- 0.15 & 0.12 & \xsecvil{5.58}{1.29}{0.62}{0.74} & & \phibi & $-2.60$ & \xsecvil{0.35}{0.13}{0.06}{0.07} \\[8pt]
0.15 -- 0.22 & 0.19 & \xsecvil{6.88}{1.31}{0.73}{0.89} & & \phibii & $-1.92$ & \xsecvil{1.00}{0.19}{0.09}{0.12} \\[8pt]
0.22 -- 0.36 & 0.29 & \xsecvil{5.07}{0.70}{0.63}{0.79} & & \phibiii & $-1.16$ & \xsecvil{1.21}{0.21}{0.11}{0.13} \\[8pt]
0.36 -- 0.50 & 0.42 & \xsecvil{3.42}{0.58}{0.40}{0.51} & & \phibiv & $-0.42$ & \xsecvil{0.38}{0.10}{0.09}{0.10} \\[8pt]
0.50 -- 0.64 & 0.57 & \xsecvil{3.86}{0.64}{0.35}{0.52} & & \phibv & $\fm0.44$ & \xsecvil{0.53}{0.12}{0.10}{0.11} \\[8pt]
             &      &                                  & &  \phibvi & $\fm1.16$ & \xsecvil{1.20}{0.19}{0.11}{0.14} \\[8pt]
0.08-0.64 & 0.33 & \xsecvil{4.65}{0.36}{0.51}{0.61}    & & \phibvii & $\fm1.93$ & \xsecvil{0.82}{0.17}{0.07}{0.09} \\[8pt]
             &      &                                  & &  \phibviii & $\fm2.67$ & \xsecvil{0.43}{0.16}{0.06}{0.08} \\[8pt]
\hline
\end{tabular}}
\end{table}

\begin{table}[p]
\caption{\label{tab:app_nu_1}
Numerical values of the cross sections shown in Fig.~\ref{Fig:Evolution_nu} for
6.4\,\text{GeV} $<  \nu  <$  8.5\,\text{GeV} with the mean values of $|t|$ and $\phi$ in each bin.
}
\centering
\resizebox{\textwidth}{!}{%
\begin{tabular}{c c c c c c c}
\hline
 &  & \hspace{1cm} & & & & \\[-4pt]
{\footnotesize $|t|$-range} & $\langle |t|\rangle$ $\bigl[\text{\GeVcsq}\bigr]$ & $\Big\langle \frac{\di \sigma}{\di |t|} \Big\rangle~\left[\frac{\text{nb}}{\text{\GeVcsq}} \right]$ & \hspace{0.5cm} & {\footnotesize $\phi$-range } & $\langle \phi \rangle$ [rad] & $\frac{1}{\Delta |t| }\Big\langle \frac{\di \sigma}{\di \phi} \Big\rangle ~\left[\frac{\text{nb}}{\text{\GeVcsq}} \right]$  \\[8pt]
\hline\\[-6pt]
0.08 -- 0.15 & 0.12 & \xsecvil{34.22}{4.22}{4.48}{5.13} & & \phibi & $-2.67$ & \xsecvil{3.84}{0.99}{0.50}{0.64} \\[8pt]
0.15 -- 0.22 & 0.18 & \xsecvil{54.80}{6.83}{5.48}{6.80} & & \phibii & $-1.93$ & \xsecvil{7.57}{1.22}{0.61}{0.79} \\[8pt]
0.22 -- 0.36 & 0.28 & \xsecvil{40.05}{3.84}{3.72}{5.25} & & \phibiii & $-1.22$ & \xsecvil{9.38}{1.14}{0.76}{0.99} \\[8pt]
0.36 -- 0.50 & 0.43 & \xsecvil{35.11}{4.81}{3.16}{4.53} & & \phibiv & $-0.51$ & \xsecvil{3.98}{0.68}{0.63}{0.77} \\[8pt]
0.50 -- 0.64 & 0.57 & \xsecvil{39.30}{6.72}{3.18}{5.11} & & \phibv & $\fm0.49$ & \xsecvil{4.16}{0.91}{0.64}{0.78} \\[8pt]
             &      &                                   & & \phibvi & $\fm1.23$ & \xsecvil{9.55}{1.18}{0.79}{1.03} \\[8pt]
0.08 -- 0.64 & 0.31 & \xsecvil{39.75}{2.49}{3.70}{4.73} & & \phibvii & $\fm1.94$ & \xsecvil{9.40}{1.79}{0.74}{0.98} \\[8pt]
             &      &                                   & & \phibviii & $\fm2.74$ & \xsecvil{2.73}{0.64}{0.37}{0.46} \\[8pt]
\hline
\end{tabular}}
\end{table}

\begin{table}[p]
\caption{\label{tab:app_nu_2}
Numerical values of the  cross sections shown in Fig.~\ref{Fig:Evolution_nu} for
8.5\,\text{GeV} $<  \nu  <$  13.9\,\text{GeV}  with the mean values of $|t|$ and $\phi$ in each bin.
}
\centering
\resizebox{\textwidth}{!}{%
\begin{tabular}{c c c c c c c}
\hline
 &  & \hspace{1cm} & & & & \\[-4pt]
{\footnotesize $|t|$-range} & $\langle |t|\rangle$ $\bigl[\text{\GeVcsq}\bigr]$ & $\Big\langle \frac{\di \sigma}{\di |t|} \Big\rangle~\left[\frac{\text{nb}}{\text{\GeVcsq}} \right]$ & \hspace{0.5cm} & {\footnotesize $\phi$-range } & $\langle \phi \rangle$ [rad] & $\frac{1}{\Delta |t| }\Big\langle \frac{\di \sigma}{\di \phi} \Big\rangle ~\left[\frac{\text{nb}}{\text{\GeVcsq}} \right]$  \\[8pt]
\hline\\[-6pt]
0.08 -- 0.15 & 0.12 & \xsecvil{19.69}{3.00}{1.87}{2.24} & & \phibi & $-2.67$ & \xsecvil{1.76}{0.49}{0.25}{0.29} \\[8pt]
0.15 -- 0.22 & 0.18 & \xsecvil{29.78}{4.48}{2.77}{3.34} & & \phibii & $-1.97$ & \xsecvil{3.93}{0.62}{0.34}{0.42} \\[8pt]
0.22 -- 0.36 & 0.29 & \xsecvil{19.63}{2.43}{2.12}{2.67} & & \phibiii & $-1.20$ & \xsecvil{3.76}{0.60}{0.32}{0.39} \\[8pt]
0.36 -- 0.50 & 0.42 & \xsecvil{10.83}{1.62}{1.34}{1.61} & & \phibiv & $-0.40$ & \xsecvil{1.49}{0.39}{0.27}{0.31} \\[8pt]
0.50 -- 0.64 & 0.57 & \xsecvil{12.66}{2.40}{1.16}{1.65} & & \phibv & $\fm0.46$ & \xsecvil{1.81}{0.37}{0.32}{0.36} \\[8pt]
             &      &                                   & & \phibvi & $\fm1.20$ & \xsecvil{3.97}{0.51}{0.33}{0.42} \\[8pt]
0.08 -- 0.64 & 0.30 & \xsecvil{17.06}{1.16}{1.74}{2.05} & & \phibvii & $\fm1.91$ & \xsecvil{3.12}{0.51}{0.27}{0.33} \\[8pt]
             &      &                                   & & \phibviii & $\fm2.69$ & \xsecvil{1.88}{0.62}{0.27}{0.30} \\[8pt]
\hline
\end{tabular}}
\end{table}

\begin{table}[p]
\caption{\label{tab:app_nu_3}
Numerical values of the cross sections shown in Fig.~\ref{Fig:Evolution_nu} for
13.9\,\text{GeV} $<  \nu  <  40.0$\,\text{GeV} with the mean values of $|t|$ and $\phi$ in each bin.}
\centering
\resizebox{\textwidth}{!}{%
\begin{tabular}{c c c c c c c}
\hline
 &  & \hspace{1cm} & & & & \\[-4pt]
{\footnotesize $|t|$-range} & $\langle |t|\rangle$ $\bigl[\text{\GeVcsq}\bigr]$ & $\Big\langle \frac{\di \sigma}{\di |t|} \Big\rangle~\left[\frac{\text{nb}}{\text{\GeVcsq}} \right]$ & \hspace{0.5cm} & {\footnotesize $\phi$-range } & $\langle \phi \rangle$ [rad] & $\frac{1}{\Delta |t| }\Big\langle \frac{\di \sigma}{\di \phi} \Big\rangle ~\left[\frac{\text{nb}}{\text{\GeVcsq}} \right]$  \\[8pt]
\hline\\[-6pt]
0.08 -- 0.15 & 0.12 & \xsecvil{5.92}{1.26}{0.73}{0.56} & & \phibi   & $-$2.56 & \xsecvil{0.11}{0.05}{0.03}{0.02}  \\[8pt]
0.15 -- 0.22 & 0.18 & \xsecvil{3.80}{0.93}{0.48}{0.38} & & \phibii  & $-$1.94 & \xsecvil{0.42}{0.12}{0.05}{0.05}  \\[8pt]
0.22 -- 0.36 & 0.27 & \xsecvil{2.06}{0.43}{0.32}{0.30} & & \phibiii & $-$1.25 & \xsecvil{0.53}{0.15}{0.06}{0.06}  \\[8pt]
0.36 -- 0.50 & 0.41 & \xsecvil{0.96}{0.30}{0.23}{0.22} & & \phibiv  & $-$0.40 & \xsecvil{0.16}{0.06}{0.04}{0.03}  \\[8pt]
0.50 -- 0.64 & 0.56 & \xsecvil{0.91}{0.29}{0.14}{0.14} & & \phibv   & \fm0.56 & \xsecvil{0.23}{0.08}{0.04}{0.04}  \\[8pt]
             &      &                                  & & \phibvi  & \fm1.20 & \xsecvil{0.73}{0.14}{0.07}{0.07}  \\[8pt]
0.08 -- 0.64 & 0.24 &  \xsecvil{2.21}{0.25}{0.31}{0.26}& & \phibvii & \fm1.89 & \xsecvil{0.41}{0.10}{0.05}{0.04}  \\[8pt]
             &      &                                  & & \phibviii& \fm2.59 & \xsecvil{0.23}{0.13}{0.06}{0.04}  \\[8pt]
\hline
\end{tabular}}
\end{table}

\begin{table}[p]
\caption{\label{tab:app_t_1}
Numerical values of the cross sections shown in Fig.~\ref{fig:xsection_phi_distr_5tbins} with the mean values of $\phi$ in each of the equally spaced bins.
}
\centering
\vspace{0.2cm}
\resizebox{\textwidth}{!}{%
\begin{tabular}{c c}
\multicolumn{2}{c}{0.08\,\text{\GeVcsq} $< |t| <$ 0.15\,\text{\GeVcsq}} \\[6pt]
\hline\\[-6pt]
$\langle \phi \rangle$ [rad] & $\frac{1}{\Delta |t| } \Big\langle \frac{\di \sigma}{  \di \phi} \Big\rangle ~\left[\frac{\text{nb}}{\text{\GeVcsq}}\right]$ \\[8pt]
\hline\\[-6pt]
$-$2.61 & \xsecvil{0.59}{0.30}{0.09}{0.10} \\[4pt]
$-$1.94 & \xsecvil{1.27}{0.32}{0.11}{0.14} \\[4pt]
$-$1.22 & \xsecvil{3.10}{0.84}{0.27}{0.33} \\[4pt]
$-$0.47 & \xsecvil{0.88}{0.33}{0.15}{0.17} \\[4pt]
\fm0.53 & \xsecvil{0.75}{0.26}{0.13}{0.15} \\[4pt]
\fm1.16 & \xsecvil{3.22}{0.58}{0.28}{0.34} \\[4pt]
\fm1.89 & \xsecvil{1.95}{0.36}{0.17}{0.21} \\[4pt]
\fm2.73 & \xsecvil{0.74}{0.59}{0.11}{0.12} \\[4pt]
\hline
\end{tabular}
\hspace{6pt}
\begin{tabular}{c c}
\multicolumn{2}{c}{0.15\,\text{\GeVcsq} $< |t| <$ 0.22\,\text{\GeVcsq}} \\[6pt]
\hline\\[-6pt]
$\langle \phi \rangle$ [rad] & $\frac{1}{\Delta |t| } \Big\langle \frac{\di \sigma}{  \di \phi} \Big\rangle ~~\left[\frac{\text{nb}}{\text{\GeVcsq}}\right]$ \\[8pt]
\hline\\[-6pt]
$-$2.70 & \xsecvil{1.23}{0.49}{0.18}{0.21} \\[4pt]
$-$1.98 & \xsecvil{3.18}{0.74}{0.28}{0.34} \\[4pt]
$-$1.19 & \xsecvil{1.69}{0.38}{0.15}{0.18} \\[4pt]
$-$0.51 & \xsecvil{1.13}{0.39}{0.19}{0.22} \\[4pt]
\fm0.52 & \xsecvil{0.56}{0.26}{0.09}{0.11} \\[4pt]
\fm1.26 & \xsecvil{2.68}{0.48}{0.23}{0.29} \\[4pt]
\fm1.98 & \xsecvil{2.31}{0.55}{0.20}{0.25} \\[4pt]
\fm2.72 & \xsecvil{1.86}{1.25}{0.29}{0.32} \\[4pt]
\hline
\end{tabular}
\hspace{6pt}
\begin{tabular}{c c}
\multicolumn{2}{c}{0.22\,\text{\GeVcsq} $< |t| <$ 0.36\,\text{\GeVcsq}} \\[6pt]
\hline\\[-6pt]
$\langle \phi \rangle$ [rad] & $\frac{1}{\Delta |t| } \Big\langle \frac{\di \sigma}{  \di \phi} \Big\rangle ~~\left[\frac{\text{nb}}{\text{\GeVcsq}}\right]$ \\[8pt]
\hline\\[-6pt]
$-$2.65 & \xsecvil{0.66}{0.22}{0.13}{0.14} \\[4pt]
$-$1.92 & \xsecvil{1.50}{0.29}{0.16}{0.19} \\[4pt]
$-$1.24 & \xsecvil{1.24}{0.25}{0.13}{0.16} \\[4pt]
$-$0.43 & \xsecvil{0.78}{0.21}{0.18}{0.20} \\[4pt]
\fm0.48 & \xsecvil{0.93}{0.26}{0.22}{0.24} \\[4pt]
\fm1.21 & \xsecvil{2.26}{0.38}{0.24}{0.28} \\[4pt]
\fm1.93 & \xsecvil{0.99}{0.21}{0.11}{0.12} \\[4pt]
\fm2.63 & \xsecvil{0.33}{0.12}{0.07}{0.07} \\[4pt]
\hline
\end{tabular}
}\\
\vspace{0.6cm}
\resizebox{0.66\textwidth}{!}{%
\begin{tabular}{c c}
\multicolumn{2}{c}{0.36\,\text{\GeVcsq} $< |t| <$ 0.50\,\text{\GeVcsq}} \\[6pt]
\hline\\[-6pt]
$\langle \phi \rangle$ [rad] & $\frac{1}{\Delta |t| } \Big\langle \frac{\di \sigma}{  \di \phi} \Big\rangle ~~\left[\frac{\text{nb}}{\text{\GeVcsq}}\right]$ \\[8pt]
\hline\\[-6pt]
$-$2.67 & \xsecvil{0.22}{0.09}{0.04}{0.05} \\[4pt]
$-$1.95 & \xsecvil{1.05}{0.30}{0.11}{0.13} \\[4pt]
$-$1.20 & \xsecvil{1.28}{0.27}{0.14}{0.16} \\[4pt]
$-$0.43 & \xsecvil{0.19}{0.10}{0.04}{0.05} \\[4pt]
\fm0.50 & \xsecvil{0.42}{0.14}{0.10}{0.11} \\[4pt]
\fm1.21 & \xsecvil{1.00}{0.21}{0.11}{0.13} \\[4pt]
\fm1.91 & \xsecvil{1.01}{0.23}{0.11}{0.13} \\[4pt]
\fm2.72 & \xsecvil{0.44}{0.18}{0.09}{0.09} \\[4pt]
\hline
\end{tabular}
\hspace{6pt}
\begin{tabular}{c c}
\multicolumn{2}{c}{0.50\,\text{\GeVcsq} $< |t| <$ 0.64\,\text{\GeVcsq}} \\[6pt]
\hline\\[-6pt]
$\langle \phi \rangle$ [rad] & $\frac{1}{\Delta |t| } \Big\langle \frac{\di \sigma}{  \di \phi} \Big\rangle ~~\left[\frac{\text{nb}}{\text{\GeVcsq}}\right]$ \\[8pt]
\hline\\[-6pt]
$-$2.66 & \xsecvil{0.35}{0.17}{0.07}{0.07} \\[4pt]
$-$1.96 & \xsecvil{0.74}{0.24}{0.08}{0.09} \\[4pt]
$-$1.22 & \xsecvil{1.18}{0.26}{0.13}{0.15} \\[4pt]
$-$0.46 & \xsecvil{0.28}{0.10}{0.07}{0.07} \\[4pt]
\fm0.42 & \xsecvil{0.68}{0.20}{0.16}{0.17} \\[4pt]
\fm1.25 & \xsecvil{0.94}{0.26}{0.10}{0.12} \\[4pt]
\fm1.88 & \xsecvil{1.39}{0.45}{0.15}{0.17} \\[4pt]
\fm2.78 & \xsecvil{0.69}{0.32}{0.13}{0.15} \\[4pt]
\hline
\end{tabular}
}
\end{table}
\clearpage

\bibliography{COMPASS_pi0_xsec_2016_data_arXiv}{}

\center{\textbf{The COMPASS Collaboration}}

\vspace{10pt}
\begin{flushleft}
G.~D.~Alexeev$^\textrm{{\footnotesize\hyperlink{hl:dubna}{28}}}$\orcidlink{0009-0007-0196-8178},
M.~G.~Alexeev$^\textrm{{\footnotesize\hyperlink{hl:turin_u}{20},\hyperlink{hl:turin_i}{19}}}$\orcidlink{0000-0002-7306-8255},
C.~Alice$^\textrm{{\footnotesize\hyperlink{hl:turin_u}{20},\hyperlink{hl:turin_i}{19}}}$\orcidlink{0000-0001-6297-9857},
A.~Amoroso$^\textrm{{\footnotesize\hyperlink{hl:turin_u}{20},\hyperlink{hl:turin_i}{19}}}$\orcidlink{0000-0002-3095-8610},
V.~Andrieux$^\textrm{{\footnotesize\hyperlink{hl:illinois}{33}}}$\orcidlink{0000-0001-9957-9910},
V.~Anosov$^\textrm{{\footnotesize\hyperlink{hl:dubna}{28}}}$\orcidlink{0009-0003-3595-9561},
K.~Augsten$^\textrm{{\footnotesize\hyperlink{hl:praguectu}{4}}}$\orcidlink{0000-0001-8324-0576},
W.~Augustyniak$^\textrm{{\footnotesize\hyperlink{hl:warsaw}{23}}}$,
C.~D.~R.~Azevedo$^\textrm{{\footnotesize\hyperlink{hl:aveiro}{26}}}$\orcidlink{0000-0002-0012-9918},
B.~Badelek$^\textrm{{\footnotesize\hyperlink{hl:warsawu}{25}}}$\orcidlink{0000-0002-4082-1466},
J.~Barth$^\textrm{{\footnotesize\hyperlink{hl:bonniskp}{8}}}$\orcidlink{0009-0003-0891-9935},
R.~Beck$^\textrm{{\footnotesize\hyperlink{hl:bonniskp}{8}}}$,
J.~Beckers$^\textrm{{\footnotesize\hyperlink{hl:munichtu}{12}}}$\orcidlink{0009-0009-7186-255X},
Y.~Bedfer$^\textrm{{\footnotesize\hyperlink{hl:saclay}{6}}}$\orcidlink{0000-0002-5198-1852},
J.~Bernhard$^\textrm{{\footnotesize\hyperlink{hl:cern}{30}}}$\orcidlink{0000-0001-9256-971X},
M.~Bodlak$^\textrm{{\footnotesize\hyperlink{hl:praguecu}{5}}}$,
F.~Bradamante$^\textrm{{\footnotesize\hyperlink{hl:triest_i}{17}}}$\orcidlink{0000-0001-6136-376X},
A.~Bressan$^\textrm{{\footnotesize\hyperlink{hl:triest_u}{18},\hyperlink{hl:triest_i}{17}}}$\orcidlink{0000-0002-3718-6377},
W.-C.~Chang$^\textrm{{\footnotesize\hyperlink{hl:taipei}{31}}}$\orcidlink{0000-0002-1695-7830},
C.~Chatterjee$^\textrm{{\footnotesize\hyperlink{hl:triest_i}{17},\hyperlink{hl:a}{a}}}$\orcidlink{0000-0001-7784-3792},
M.~Chiosso$^\textrm{{\footnotesize\hyperlink{hl:turin_u}{20},\hyperlink{hl:turin_i}{19}}}$\orcidlink{0000-0001-6994-8551},
S.-U.~Chung$^\textrm{{\footnotesize\hyperlink{hl:munichtu}{12},\hyperlink{hl:i}{i},\hyperlink{hl:i1}{i1}}}$,
A.~Cicuttin$^\textrm{{\footnotesize\hyperlink{hl:triest_i}{17},\hyperlink{hl:triest_a}{16}}}$\orcidlink{0000-0002-3645-9791},
P.~M.~M.~Correia$^\textrm{{\footnotesize\hyperlink{hl:aveiro}{26}}}$\orcidlink{0000-0001-7292-7735},
M.~L.~Crespo$^\textrm{{\footnotesize\hyperlink{hl:triest_i}{17},\hyperlink{hl:triest_a}{16}}}$\orcidlink{0000-0002-5483-3388}
% ...

\vspace{10pt}
\hypertarget{hl:aanl}{$^\textrm{{\footnotesize 1}}$\footnotesize~A.I. Alikhanyan National Science Laboratory, 2 Alikhanyan Br. Street, 0036, Yerevan, Armenia$^\textrm{{\tiny\hyperlink{hl:A}{A}}}$\\}
\hypertarget{hl:brno}{$^\textrm{{\footnotesize 2}}$\footnotesize~Institute of Scientific Instruments of the CAS, 61264 Brno, Czech Republic$^\textrm{{\tiny\hyperlink{hl:B}{B}}}$\\}
\hypertarget{hl:liberec}{$^\textrm{{\footnotesize 3}}$\footnotesize~Technical University in Liberec, 46117 Liberec, Czech Republic$^\textrm{{\tiny\hyperlink{hl:B}{B}}}$\\}
\hypertarget{hl:praguectu}{$^\textrm{{\footnotesize 4}}$\footnotesize~Czech Technical University in Prague, 16636 Prague, Czech Republic$^\textrm{{\tiny\hyperlink{hl:B}{B}}}$\\}
\hypertarget{hl:praguecu}{$^\textrm{{\footnotesize 5}}$\footnotesize~Charles University, Faculty of Mathematics and Physics, 12116 Prague, Czech Republic$^\textrm{{\tiny\hyperlink{hl:B}{B}}}$\\}
\hypertarget{hl:saclay}{$^\textrm{{\footnotesize 6}}$\footnotesize~IRFU, CEA, Universit\'e Paris-Saclay, 91191 Gif-sur-Yvette, France\\}
\hypertarget{hl:bochum}{$^\textrm{{\footnotesize 7}}$\footnotesize~Universit\"at Bochum, Institut f\"ur Experimentalphysik, 44780 Bochum, Germany$^\textrm{{\tiny\hyperlink{hl:C}{C}}}$\\}
\hypertarget{hl:bonniskp}{$^\textrm{{\footnotesize 8}}$\footnotesize~Universit\"at Bonn, Helmholtz-Institut f\"ur  Strahlen- und Kernphysik, 53115 Bonn, Germany$^\textrm{{\tiny\hyperlink{hl:C}{C}}}$\\}
\hypertarget{hl:bonnpi}{$^\textrm{{\footnotesize 9}}$\footnotesize~Universit\"at Bonn, Physikalisches Institut, 53115 Bonn, Germany$^\textrm{{\tiny\hyperlink{hl:C}{C}}}$\\}
\hypertarget{hl:freiburg}{$^\textrm{{\footnotesize 10}}$\footnotesize~Universit\"at Freiburg, Physikalisches Institut, 79104 Freiburg, Germany$^\textrm{{\tiny\hyperlink{hl:C}{C}}}$\\}
\hypertarget{hl:mainz}{$^\textrm{{\footnotesize 11}}$\footnotesize~Universit\"at Mainz, Institut f\"ur Kernphysik, 55099 Mainz, Germany$^\textrm{{\tiny\hyperlink{hl:C}{C}}}$\\}
\hypertarget{hl:munichtu}{$^\textrm{{\footnotesize 12}}$\footnotesize~Technische Universit\"at M\"unchen, Physik Dept., 85748 Garching, Germany$^\textrm{{\tiny\hyperlink{hl:C}{C}}}$\\}
\hypertarget{hl:munichuni}{$^\textrm{{\footnotesize 13}}$\footnotesize~Ludwig-Maximilians-Universit\"at, 80539 M\"unchen, Germany\\}
\hypertarget{hl:calcutta}{$^\textrm{{\footnotesize 14}}$\footnotesize~Matrivani Institute of Experimental Research \& Education, Calcutta-700 030, India$^\textrm{{\tiny\hyperlink{hl:D}{D}}}$\\}
\hypertarget{hl:telaviv}{$^\textrm{{\footnotesize 15}}$\footnotesize~Tel Aviv University, School of Physics and Astronomy, 69978 Tel Aviv, Israel$^\textrm{{\tiny\hyperlink{hl:E}{E}}}$\\}
\hypertarget{hl:triest_a}{$^\textrm{{\footnotesize 16}}$\footnotesize~Abdus Salam ICTP, 34151 Trieste, Italy\\}
\hypertarget{hl:triest_i}{$^\textrm{{\footnotesize 17}}$\footnotesize~Trieste Section of INFN, 34127 Trieste, Italy\\}
\hypertarget{hl:triest_u}{$^\textrm{{\footnotesize 18}}$\footnotesize~University of Trieste, Dept.\ of Physics, 34127 Trieste, Italy\\}
\hypertarget{hl:turin_i}{$^\textrm{{\footnotesize 19}}$\footnotesize~Torino Section of INFN, 10125 Torino, Italy\\}
\hypertarget{hl:turin_u}{$^\textrm{{\footnotesize 20}}$\footnotesize~University of Torino, Dept.\ of Physics, 10125 Torino, Italy\\}
\hypertarget{hl:miyazaki}{$^\textrm{{\footnotesize 21}}$\footnotesize~University of Miyazaki, Miyazaki 889-2192, Japan$^\textrm{{\tiny\hyperlink{hl:F}{F}}}$\\}
\hypertarget{hl:yamagata}{$^\textrm{{\footnotesize 22}}$\footnotesize~Yamagata University, Yamagata 992-8510, Japan$^\textrm{{\tiny\hyperlink{hl:F}{F}}}$\\}
\hypertarget{hl:warsaw}{$^\textrm{{\footnotesize 23}}$\footnotesize~National Centre for Nuclear Research, 02-093 Warsaw, Poland$^\textrm{{\tiny\hyperlink{hl:G}{G}}}$\\}
\hypertarget{hl:warsawtu}{$^\textrm{{\footnotesize 24}}$\footnotesize~Warsaw University of Technology, Institute of Radioelectronics, 00-665 Warsaw, Poland$^\textrm{{\tiny\hyperlink{hl:G}{G}}}$\\}
\hypertarget{hl:warsawu}{$^\textrm{{\footnotesize 25}}$\footnotesize~University of Warsaw, Faculty of Physics, 02-093 Warsaw, Poland$^\textrm{{\tiny\hyperlink{hl:G}{G}}}$\\}
\hypertarget{hl:aveiro}{$^\textrm{{\footnotesize 26}}$\footnotesize~University of Aveiro, I3N, Dept. of Physics, 3810-193 Aveiro, Portugal$^\textrm{{\tiny\hyperlink{hl:H}{H}}}$\\}
\hypertarget{hl:lisbon}{$^\textrm{{\footnotesize 27}}$\footnotesize~LIP, 1649-003 Lisbon, Portugal$^\textrm{{\tiny\hyperlink{hl:H}{H}}}$\\}
\hypertarget{hl:dubna}{$^\textrm{{\footnotesize 28}}$\footnotesize~Affiliated with an international laboratory covered by a cooperation agreement with CERN\\}
\hypertarget{hl:russia}{$^\textrm{{\footnotesize 29}}$\footnotesize~Affiliated with an institute covered by a cooperation agreement with CERN.\\}
\hypertarget{hl:cern}{$^\textrm{{\footnotesize 30}}$\footnotesize~CERN, 1211 Geneva 23, Switzerland\\}
\hypertarget{hl:taipei}{$^\textrm{{\footnotesize 31}}$\footnotesize~Academia Sinica, Institute of Physics, Taipei 11529, Taiwan$^\textrm{{\tiny\hyperlink{hl:I}{I}}}$\\}
\hypertarget{hl:taipeincu}{$^\textrm{{\footnotesize 32}}$\footnotesize~Center for High Energy and High Field Physics and Dept.\ of Physics, National Central University, 300 Zhongda Rd., Zhongli 320317, Taiwan$^\textrm{{\tiny\hyperlink{hl:I}{I}}}$\\}
\hypertarget{hl:illinois}{$^\textrm{{\footnotesize 33}}$\footnotesize~University of Illinois at Urbana-Champaign, Dept.\ of Physics, Urbana, IL 61801-3080, USA$^\textrm{{\tiny\hyperlink{hl:J}{J}}}$\\}

\vspace{10pt}
\hypertarget{hl:*}{$^\textrm{{\footnotesize *}}$\footnotesize~Corresponding author\\}
\hypertarget{hl:a}{$^\textrm{{\footnotesize a}}$\footnotesize~Supported by the European Union’s Horizon 2020 research and innovation programme under grant agreement STRONG–2020 - No 824093\\}
\hypertarget{hl:b}{$^\textrm{{\footnotesize b}}$\footnotesize~Retired from Ludwig-Maximilians-Universit\"at, 80539 M\"unchen, Germany\\}
\hypertarget{hl:b1}{$^\textrm{{\footnotesize b1}}$\footnotesize~Supported by the DFG cluster of excellence `Origin and Structure of the Universe' (www.universe-cluster.de) (Germany)\\}
\hypertarget{hl:c}{$^\textrm{{\footnotesize c}}$\footnotesize~Also at ORIGINS Excellence Cluster, 85748 Garching, Germany\\}
\hypertarget{hl:d}{$^\textrm{{\footnotesize d}}$\footnotesize~Also at Institut f\"ur Theoretische Physik, Universit\"at T\"ubingen, 72076 T\"ubingen, Germany\\}
\hypertarget{hl:e}{$^\textrm{{\footnotesize e}}$\footnotesize~Present address: NISER, Centre for Medical and Radiation Physics, Bubaneswar, India\\}
\hypertarget{hl:f}{$^\textrm{{\footnotesize f}}$\footnotesize~Also at University of Eastern Piedmont, 15100 Alessandria, Italy\\}
\hypertarget{hl:f1}{$^\textrm{{\footnotesize f1}}$\footnotesize~Supported by the Funds for Research 2019-22 of the University of Eastern Piedmont\\}
\hypertarget{hl:g}{$^\textrm{{\footnotesize g}}$\footnotesize~Also at Chubu University, Kasugai, Aichi 487-8501, Japan\\}
\hypertarget{hl:h}{$^\textrm{{\footnotesize h}}$\footnotesize~Also at KEK, 1-1 Oho, Tsukuba, Ibaraki 305-0801, Japan\\}
\hypertarget{hl:i}{$^\textrm{{\footnotesize i}}$\footnotesize~Also at Dept.\ of Physics, Pusan National University, Busan 609-735, Republic of Korea\\}
\hypertarget{hl:i1}{$^\textrm{{\footnotesize i1}}$\footnotesize~Also at Physics Dept., Brookhaven National Laboratory, Upton, NY 11973, USA\\}
\hypertarget{hl:j}{$^\textrm{{\footnotesize j}}$\footnotesize~Also at Fairmont State University, Department of Natural Sciences, 1201 Locust Ave, Fairmont, West Virginia 26554, USA\\}
\hypertarget{hl:k}{$^\textrm{{\footnotesize k}}$\footnotesize~Also at Dept.\ of Physics, National Kaohsiung Normal University, Kaohsiung County 824, Taiwan\\}
\hypertarget{hl:$\dagger$}{$^\textrm{{\footnotesize $\dagger$}}$\footnotesize~Deceased\\}
\hypertarget{hl:l}{$^\textrm{{\footnotesize l}}$\footnotesize~Supported by the FORTE project CZ.02.01.01/00/22 co-funded by the European Union and Ministry of Education, Youth and Sports, by the Martina Roeselov\'a Memorial Fellowship grant by the IOCB Tech Foundation and by the SGS24/149/OHK4/3T/14 project, Czech Republic\\}
\hypertarget{hl:m}{$^\textrm{{\footnotesize m}}$\footnotesize~Supported by the grant GAUK 60121, issued by Charles University, Faculty of Mathematics and Physics, and the grant SVV No. 260576, Czech Republic\\}
\hypertarget{hl:n}{$^\textrm{{\footnotesize n}}$\footnotesize~Also at RIKEN Nishina Center for Accelerator-Based Science, Wako, Saitama 351-0198, Japan\\}

\vspace{10pt}
\hypertarget{hl:A}{$^\textrm{{\tiny A}}$\footnotesize~Supported by the Higher Education and Science Committee of the Republic of Armenia (Armenia)\\}
\hypertarget{hl:B}{$^\textrm{{\tiny B}}$\footnotesize~Supported by MEYS Grants LM2018104, LM2023040 and LTT17018 and Charles University grants PRIMUS/22/SCI/017 and GAUK60121 (Czech Republic)\\}
\hypertarget{hl:C}{$^\textrm{{\tiny C}}$\footnotesize~Supported by BMBF - Bundesministerium f\"ur Bildung und Forschung (Germany)\\}
\hypertarget{hl:D}{$^\textrm{{\tiny D}}$\footnotesize~Supported by B. Sen fund (India)\\}
\hypertarget{hl:E}{$^\textrm{{\tiny E}}$\footnotesize~Supported by the Israel Academy of Sciences and Humanities (Israel)\\}
\hypertarget{hl:F}{$^\textrm{{\tiny F}}$\footnotesize~Supported by MEXT and JSPS, Grants 18002006, 20540299, 18540281 and 26247032, the Daiko and Yamada Foundations (Japan)\\}
\hypertarget{hl:G}{$^\textrm{{\tiny G}}$\footnotesize~Supported by NCN, Grant 2020/37/B/ST2/01547 (Poland)\\}
\hypertarget{hl:H}{$^\textrm{{\tiny H}}$\footnotesize~Supported by FCT, Grants DOI 10.54499/CERN/FIS-PAR/0022/2019 and DOI 10.54499/CERN/FIS-PAR/0016/2021 (Portugal)\\}
\hypertarget{hl:I}{$^\textrm{{\tiny I}}$\footnotesize~Supported by the Ministry of Science and Technology (Taiwan)\\}
\hypertarget{hl:J}{$^\textrm{{\tiny J}}$\footnotesize~Supported by the National Science Foundation, Grant no. PHY-1506416 (USA)\\}

\end{flushleft} 

\end{document}